\documentclass[a4paper,usenatbib]{mn2e}

\usepackage[utf8]{inputenc}
\usepackage[T1]{fontenc}
\usepackage{lmodern} 

\usepackage{graphicx}   
\usepackage{amsmath}    
\usepackage{amssymb}    

\newcommand{\unit}[1]{\, \mathrm{#1}}	
\newcommand{\sscript}[1]{_\mathrm{#1}}		

\title[Infalling Young Clusters in the Galactic Centre: implications for IMBHs and young stellar populations]{Infalling Young Clusters in the Galactic Centre: implications for IMBHs and young stellar populations}
\author[J. A. Petts, A. Gualandris]{J. A. Petts$^{1}$\thanks{E-mail:
j.petts@surrey.ac.uk}, A. Gualandris\\
$^{1}$University of Surrey, Guildford, United Kingdom}
\begin{document}

\date{Submitted TEMP}

\pagerange{\pageref{firstpage}--\pageref{lastpage}} \pubyear{2015}

\maketitle

\label{firstpage}

\begin{abstract}

The central parsec of the Milky Way hosts two puzzlingly young stellar populations, a tight isotropic distribution of B stars around SgrA* (the S-stars) and a disk of OB stars extending to $\sim 0.5\unit{pc}$. Using a modified version of Sverre Aarseth's direct summation code NBODY6 we explore the scenario in which a young star cluster migrates to the Galactic Centre within the lifetime of the OB disk population via dynamical friction. We find that star clusters massive and dense enough to reach the central parsec form a very massive star via physical collisions on a mass segregation timescale. We follow the evolution of the merger product using the most up to date, yet conservative, mass loss recipes for very massive stars. Over a large range of initial conditions, we find that the very massive star expels most of its mass via a strong stellar wind, eventually collapsing to form a black hole of mass $\sim 20 - 400 \unit{M\sscript{\odot}}$, incapable of bringing massive stars to the Galactic Centre. No massive intermediate mass black hole can form in this scenario. The presence of a star cluster in the central $\sim 10\unit{pc}$ within the last $15 \unit{Myr}$ would also leave a $\sim 2 \unit{pc}$ ring of massive stars, which is not currently observed. Thus, we conclude that the star cluster migration model is highly unlikely to be the origin of either young population, and in-situ formation models or binary disruptions are favoured.
\end{abstract}


\section{Introduction}

\label{Introduction.ch}

The central parsec of the Milky Way hosts almost two dozen He-1 emission-line stars \citep{Krabbe95,Paumard01} and a population of many other OB stars in a thin clockwise disk extending from $\sim0.04-1.0\unit{pc}$ \citep{Eckart99}. \citet{Feldmeier15} recently extended the range of observations up to $\sim 4\unit{pc^2}$ centred on SgrA*; the radio source associated with the supermassive black hole (SMBH) at the centre of the Milky Way. The authors show that the OB population is very centrally concentrated, with 90\% projected within the central $0.5\unit{pc}$. The clockwise disk exhibits a top heavy mass function ($\alpha \sim 1.7$, \citet{Lu13}). \citet{Krabbe95} estimate the He-1 stars to be only $\sim3-7 \unit{Myr}$ old, which is puzzling as the tremendous tidal forces in this region make it difficult for a giant molecular cloud (GMC) to remain bound long enough for gas to cool and fragment \citep{Phinney89,Morris93,Genzel03,Levin03}.

There appears to be very few He-1 stars farther than the central parsec, other than inside/near the young Arches \citep{Nagata95,Cotera96,Figer99} and Quintuplet \citep{Okuda90,Nagata90,Glass90,Figer99} clusters at $\sim 30 \unit{pc}$. This led \citet{Gerhard01} to postulate that efficient dynamical friction on star clusters forming a few parsecs from SgrA*, where GMCs can more easily cool and fragment, could bring a dense core of massive stars to the central parsec within the age of the He-1 population.

Another model suggests that in-situ formation of the clockwise disk is possible if a tidally disrupted GMC spirals in to form a small gaseous disk, which can be dense enough to become Jeans unstable and fragment into stars \citep{Bonnell08,Mapelli08,Alig11,Mapelli12,Alig13}. The infalling cloud needs to be $\sim 10^5 \unit{M\sscript{\odot}}$ in order to reproduce observations \citep{Mapelli12}. Two large gas clouds of mass $\sim5\times10^5\unit{M\sscript{\odot}}$, M-0.02-0.07 and M-0.13-0.08, are seen projected at $\sim 7$ and $\sim 13 \unit{pc}$ from the Galactic Centre, respectively \citep{Solomon72}. The top heavy mass function can
be reproduced by the in-situ model so long as the gas has a temperature greater than $100\unit{K}$, consistent with observations of the Galactic clouds. The rotation axis of the clockwise disk shows a strong transition from the inner to outer edge \citep{Lu09,Bartko09,Lu13}, suggesting that the disk is either strongly warped, or is comprised of a series of stellar streamers with significant variation in their orbital planes \citep{Bartko09}. In-situ formation is currently favoured for the clockwise disk, as an infalling cluster would likely form a disk with a constant rotation axis \citep{Perets10}. A caveat of in-situ formation is that it requires near radial orbits incident upon SgrA*, perhaps requiring cloud-cloud collisions \citep{Wardle08,Hobbs09,Alig11}. 

Interior to the disk lies a more enigmatic population of B-stars in a spatially isotropic distribution around SgrA*, with a distribution of eccentricities skewed slightly higher than a thermal distribution (\citet{Gillessen09}, \citet{Mapelli15}). These ``S-stars'' have semi-major axes less than 0.04 pc, with S0-102 having the shortest period of just $11.5\pm0.3 \unit{yrs}$, and a pericentre approach of just $\sim260 \unit{AU}$ \citep{Meyer12}. The S-star population could potentially be older than the disk population, as the brightest star in this population, S0-2, is a main sequence B0-B2.5 V star with an age less than $15 \unit{Myr}$ \citep{Martins08}. The other stars in this population have spectra consistent with main sequence stars \citep{Eisenhauer05}, and observational limits require them to be less than $20 \unit{Myr}$ old in order to be visible.

The tidal forces in this region prohibit standard star formation, so the S-stars must have formed farther out and migrated inwards. A possible formation mechanism of the S-stars is from the tidal disruption of binaries scattered to low angular momentum orbits, producing an S-star and a hyper-velocity star via the Hills mechanism \citep{Hills91}. The captured stars would have initial eccentricities greater than $0.97$ \citep{Miller05,Bromley06}, but the presence of a cusp of stellar mass black holes around SgrA* could efficiently reduce the eccentricities of these orbits via resonant relaxation within the lifetime of the stars \citep{Perets09}. Additionally, \citet{Antonini10} show that if a binary is not tidally disrupted at first pericentre passage, the Kozai-Lidov (KL) resonance \citep{Kozai62,Lidov62} can cause the binary to coalesce after a few orbital periods, producing an S-star and no hyper velocity star.

Alternatively, \citet{Chen14} show that stars from the clockwise disk can be brought very close to SgrA* via global KL like resonances, if the clockwise disk of gas originally extended down to $\sim 10^{-6} \unit{pc}$ (the lowest stable circular orbit around SgrA*). The authors also show that O/WR stars would be tidally disrupted within the region of the observed S-star cluster due to their large stellar radii, whereas B-stars could survive, in agreement with observations. Recently, \citet{Subr16} showed that a clockwise disk with $100\%$ primordial binarity can produce $\sim 20$ S-stars in less than $4\unit{Myr}$. KL oscillations can efficiently drag binaries close to SgrA*, producing an S-star and a hyper-velocity star. This mechanism produces S-stars with eccentricities lower than from the disruption of binaries originating from outside the disk. However, in order to thermalize the S-stars, $\sim 500 \unit{M\sscript{\odot}}$ in dark remnants are still required around SgrA* in order to match observations, consistent with Fokker-Planck models \citep{Hopman06}. Three confirmed eclipsing binaries are observed within the clockwise disk, all being very massive O/WR binaries \citep{Ott99,Martins06,Pfuhl14}. \cite{Pfuhl14} estimate the present day binary fraction of the disk to be $0.3^{+0.34}_{-0.21}$ at $95\%$ confidence, with a fraction greater than $0.85$ ruled out at $99\%$ confidence. More recently, \cite{Gautam16} predict that the binary fraction must be greater than $32\%$ at $90\%$ confidence.

An additional popular scenario is the transport of stars from young dense star clusters that migrate to the Galactic Centre via dynamical friction, with the aid of an intermediate mass black hole (IMBH). \citet{Kim03} showed that to survive to the central parsec from a distance $\geq 10\unit{pc}$, clusters either need to be very massive ($\sim10^6\unit{M\sscript{\odot}}$) or very dense (central density, $\rho\sscript{c} \sim 10^8 \unit{M\sscript{\odot} pc^{-3}}$). \citet{Kim04} showed that including an IMBH in the cluster means the the core density can be lowered, but only if the IMBH contains $\sim 10 \%$ of the mass of the entire cluster, far greater than is expected from runaway collisions \citep{Zwart02}.

\citet{Fujii09} (hereafter F09) revisited this problem using the tree-direct hybrid code, BRIDGE  \citep{Fujii07}, allowing the internal dynamics of the star clusters to be resolved. The small tidal limits imposed by SgrA* meant the clusters had core densities greater than $10^7 \unit{M\sscript{\odot}pc^{-3}}$, leading to runaway collisions on a mass segregation timescale \citep{Zwart02,Zwart04}. During collisions, the resulting very massive star (VMS) was rejuvenated using the formalism of \citet{Meurs89}, and collapsed to an IMBH at the end of its main sequence lifetime, extrapolated from the results of \citet{Belkus07}. The authors found that by allowing the formation of a $3-16 \times 10^3 \unit{M\sscript{\odot}}$ IMBH \citep[see also][]{Fujii10}, some stars could be carried very close to SgrA* via a 1:1 mean resonance with the infalling IMBH. The orbits of these ``Trojan stars'' were randomised by 3-body interactions with the SMBH and IMBH, constructing a spatially isotropic S-star cluster. F09's simulation ``LD64k'' transported 23 stars to the central $0.1\unit{pc}$, however, the resolution of the simulation is $\sim 0.2\unit{pc}$, set by the force softening of SgrA*. The simulation also brought 354 stars within $0.5\unit{pc}$ of SgrA*, 16 being more massive than $20\unit{M\sscript{\odot}}$, analogous to clockwise disk stars. The IMBH formed in LD64k is more massive than the observational upper limit of $\sim 10^4 \unit{M\sscript{\odot}}$, derived from VLBA measurements of SgrA* \citep{Reid04}. However, \citet{Fujii10} state that an IMBH of $1500 \unit{M\sscript{\odot}}$ is sufficient for the randomisation of stars \citep[see also][]{Merritt09}.

Despite the successes of the F09 model, IMBH formation in young dense star clusters may be prohibited. VMSs of the order $10^3 \unit{M\sscript{\odot}}$ are expected to have luminosities greater than $10^7 L\sscript{\odot}$ \citep{Kudritzki02,Nadyozhin05,Belkus07}, driving strong stellar winds. F09 assumed the mass loss rate of stars more massive than $300\unit{M\sscript{\odot}}$ to be linear with mass, however, recent work on VMS winds show steeper relations for stars that approach the Eddington limit \citep{Kudritzki02,Vink06,Vink11}. F09's model also neglected the effect of the evolving chemical composition on the luminosity, and hence the mass loss, of the VMS \citep{Nadyozhin05}. We note that the initial mass function (IMF) used in F09, although employed due to numerical constraints, meant there were ten times more massive stars than expected from a full Kroupa IMF, leading to an increased collision rate and buildup of the VMS mass.

No conclusive evidence for the existence of IMBHs in star clusters has yet been found \citep[See][for a comprehensive review on IMBHs in globular clusters]{Lutzgendorf13,Lutzgendorf15}. Sufficiently high mass loss could cause VMSs to end their lives as stellar mass black holes or pair-instability supernovae at low metallicity \citep{Heger02}. Pair-instability supernovae candidates have recently been found at metallicities as high as $\sim 0.1 Z\sscript{\odot}$ \citep{GalYam09,Cooke12}, with expected progenitors of several hundred solar masses \citep{Chen15}.

The most massive star observed, R136a1, is a $265^{+80}_{-35} \unit{M\sscript{\odot}}$ star in the 30 Doradus region of the Large Magellanic Cloud (LMC) \citep{Crowther10}, with metallicity $Z = 0.43 Z\sscript{\odot}$. \citet{Crowther10} suggest that it could be a very rare main sequence star, with a zero age main sequence mass of $320^{+100}_{-40} \unit{M\sscript{\odot}}$. However, it could be the collision product of a few massive stars. R136a1 has a large inferred mass loss rate of $(5.1^{+0.9}_{-0.8})\times 10^{-5} \unit{M\sscript{\odot}{yr}^{-1}}$, $\sim 0.1$ dex larger than the theoretical predictions of \citet{Vink01}. \citet{Belkus07} predict that the evolution of all stars more massive than $300 \unit{M\sscript{\odot}}$ is dominated by stellar winds, with similar lifetimes of $\sim2-3\unit{Myr}$. As such, it is not surprising that R136a1 is the most massive star currently observed, as more massive VMSs should be rare and short lived.

Whilst it may be unlikely for an IMBH to form at solar metallicity, a VMS could transport stars to SgrA* within its lifetime. In this paper we test the feasibility of the star cluster migration scenario as the origin of either young population in the Galactic Centre.

We evolve direct N-body models of star clusters in the Galactic Centre, using the GPU-accelerated code NBODY6df, a modified version of Sverre Aarseth's NBODY6 \citep{Aarseth1999,Nitadori12} which includes the effects of dynamical friction semi-analytically \citep{Petts15,Petts16}. In section \S \ref{theory.ch} we describe the theory behind our dynamical friction and stellar evolution models. In section \S \ref{Numerical_Method.ch} we describe the numerical implementation. Section \S \ref{IC.ch} discusses prior constraints on the initial conditions and describes the parameters of the simulations performed. In sections \S \ref{results.ch} and \S \ref{discussion.ch}, we present our results and discuss their implications for the origin of the young populations. Finally, we present our conclusions in section \S \ref{Conclusions.ch}.

\section[]{Theory}
\label{theory.ch}

\subsection{Dynamical friction}

The dynamical friction model used in this paper is a semi-analytic implementation of Chandrasekhar's dynamical friction \citep{Chandrasekhar43}, described in \citet{Petts15}, which provides an accurate description of the drag force on star clusters orbiting in analytic spherical background distributions of asymptotic inner slope $\gamma = 0.5,..,3$ \citep[see also][for a generalised model accurate also for the cored $\gamma=0$ case]{Petts16}. The novelty of our model is the use of physically motivated, radially varying maximum and minimum impact parameters ($b\sscript{max}$ and $b\sscript{min}$ respectively), which vary based on the local properties of the background. The dynamical friction force is given by:

\begin{equation}
	\frac{d\bmath{v}\sscript{cl}}{dt} = -4\pi G^2 M\sscript{cl} \rho \log(\Lambda) f(v\sscript{*} < v\sscript{cl}) \frac{\bmath{v}\sscript{cl}}{{v^3\sscript{cl}}}
	\label{dynfric.eq}
\end{equation}
where $\bmath{v}\sscript{cl}$ is the cluster velocity, $M\sscript{cl}$ is the cluster mass, $\rho$ is the local background density and $f(v\sscript{*} < v\sscript{cl})$ is the fraction of stars moving slower than the cluster; assuming a Maxwellian distribution of velocities, valid in the cuspy models explored here. The Coulomb logarithm is given by:

\begin{equation}
	\log(\Lambda) = \log\left(\frac{b\sscript{max}}{b\sscript{min}}\right) = \log\left(\frac{\mathrm{min}(\rho(R\sscript{g})/|\nabla\rho(R\sscript{g})|,R\sscript{g})}{\mathrm{max}\left(r\sscript{hm}, GM\sscript{cl}/v\sscript{cl}^2\right)}\right),
	\label{coulog.eq}
\end{equation}
where $R\sscript{g}$ is the galactocentric distance of the cluster and $r\sscript{hm}$ is the half mass radius of the cluster. When coupled with the $N$-body dynamics, $r\sscript{hm}$ is the live half mass radius, and  $M\sscript{cl}$ is well represented by the cluster mass enclosed within its tidal radius, including stars with energies above the escape energy.

\subsection{Evolution of very massive stars}
\label{VMS_evo.ch}

\citet{Nadyozhin05} present similarity theory models of VMSs, for which the stellar properties can be calculated by solving a set of differential equations \citep{Imshennik68}. VMSs are predicted to have large convective cores containing more than $85\%$ of the mass, surrounded by a thin extensive radiative envelope. In such stars the opacity becomes larger than the electron scattering value, and can be considered to come from Thomson scattering alone. Utilising such approximations, the authors  provide simple formulae to calculate the core mass and luminosity, as functions of stellar mass and chemical composition.

The luminosity of stars with $\mu^2M \geq 100$ can be found by substituting equation 36 of \citet{Nadyozhin05} into their equation 34:

\begin{equation}
	L \approx \frac{64826 M \left(1 - 4.5/\sqrt{\mu^2 M} \right)}{1 + X},
	\label{Lum.eq}
\end{equation}
where $L$ is the luminosity, $M$ is the mass of the VMS, $X$ is the core hydrogen abundance, and $\mu$ is the mean atomic mass of the core. Assuming a fully ionised plasma, $\mu$ takes the form:
\begin{equation}
	\mu = \frac{4}{6X + Y + 2},
	\label{mu.eq}
\end{equation}
where $Y$ is the core helium abundance. Equation \ref{Lum.eq} shows that at very large masses $L \propto M$. However, unlike the F09 model, this formulation of the luminosity explicitly includes an $L \propto (1+X)^{-1}$ dependence. As the mass loss rate depends on $L$, this leads to an increased mass loss rate in the late stages of evolution (see section \S \ref{mass_loss.ch}).

\citet{Belkus07} (hereafter B07) modelled the evolution of VMSs with zero age main sequence (ZAMS) masses of up to $1000M\sscript{\odot}$, assumed to have formed via runaway collisions in a young dense star cluster. The authors numerically evolve the chemical composition of the star through the Core Hydrogen Burning (CHB) and Core Helium Burning (CHeB) phases via conservation of energy and mass loss from the stellar wind. In this section we briefly outline the model of \citet{Belkus07} and describe how we include stellar collisions and their effect on VMS evolution.

As VMSs have large convective cores, one can reasonably approximate them as homogeneous (verified to be a good approximation down to $120 \unit{M\sscript{\odot}}$, B07). Applying conservation of energy, the hydrogen fraction in the core during CHB evolves as equation 1 of B07:

\begin{equation}
	M\sscript{cc}(\mu,M)\frac{dX}{dt} = - \frac{L(\mu,M)}{\epsilon\sscript{H}},
	\label{CHB.eq}
\end{equation}
where $M\sscript{cc}$ is the mass of the convective core and $\epsilon\sscript{H}$ is the hydrogen burning efficiency (i.e. the energy released by fusing one mass unit of hydrogen to helium).

When the core is depleted of hydrogen, the VMS burns helium via equation 4 of B07 (see also \citet{Langer89b}):

\begin{subequations}
\begin{align}
	M\sscript{cc}(\mu,M) \frac{dY}{dt} = -\frac{L(\mu,M)}{\epsilon\sscript{ratio}},\\
	\epsilon\sscript{ratio} = \left[\left(\frac{B\sscript{Y}}{A\sscript{Y}} - \frac{B\sscript{O}}{A\sscript{O}}\right)+\left(\frac{B\sscript{C}}{A\sscript{C}} - \frac{B\sscript{O}}{A\sscript{O}}\right) C'(Y)\right],
	\label{CHeB.eq}
\end{align}
\end{subequations}
where $\epsilon\sscript{ratio}$ accounts for the fact that C and O are produced in a non-constant ratio, affecting the energy production per unit mass of helium burnt. Here, $A$ and $B$ are the atomic weights and binding energy of nuclei; with subscripts $Y$, $C$ and $O$ representing helium, carbon and oxygen respectively. $C'(Y)$ is the derivative of the $C(Y)$ fit from \citet{Langer89} with respect to $Y$ (see B07 for the derivation of Equation \ref{CHeB.eq}). During CHeB, $\mu$ is defined as \citep{Nadyozhin05}:
\begin{subequations}
\begin{align}
	\mu = \frac{48}{36Y + 28C + 27O},\\
	\intertext{which by assuming $Y+C+O=1$ and using the fit to $C(Y)$ by \citet{Langer89}, can be rewritten solely as a function of $Y$ as:}
	\mu = \frac{48}{19Y + C(Y) + 27}.\label{mu_WR.eq}
\end{align}
\end{subequations}
Subsequent stages of evolution are rapid and explosive. We assume that after core helium burning the remnant collapses to a black hole with no significant mass loss.

\subsubsection{Mass loss}
\label{mass_loss.ch}
The chemical evolution of the VMS is coupled to the mass evolution, as the luminosity of the star sets the wind strength. \citet{Vink11} (hereafter V11) show that the wind strength is heavily dependent on the proximity to the Eddington limit, when gravity is completely counterbalanced by the radiative forces, i.e. $g\sscript{rad}/g\sscript{grav} = 1$, where $g\sscript{rad}$ and $g\sscript{grav}$ are the radiative and gravitational forces, respectively. For a fully ionised plasma, the Eddington parameter, $\Gamma\sscript{e}$, is dominated by free electrons and is approximately constant throughout the star (V11):

\begin{equation}
	\Gamma\sscript{e} = \frac{g\sscript{rad}}{g\sscript{grav}} = 10^{-4.813} (1 + X\sscript{s}) \left(\frac{L}{L\sscript{\odot}}\right) \left(\frac{M}{M\sscript{\odot}}\right)^{-1},
\label{Edd_para.eq}
\end{equation}
where $X\sscript{s}$ is the surface hydrogen abundance of the star. V11's fig. 2 shows that the logarithmic difference between the empirical \citet{Vink01} (here after V01) rates and the VMS rates follow a tight relation with $\Gamma\sscript{e}$, almost independent of mass. The authors find that the mass loss rate is proportional to:

\begin{equation}
	\dot{M} \propto
	\begin{cases}
            \Gamma\sscript{e}^{2.2}, &         \text{if } 0.4< \Gamma\sscript{e} < 0.7\\
            \Gamma\sscript{e}^{4.77}, &         \text{if } 0.7<\Gamma\sscript{e} < 0.95.
    \end{cases}
\end{equation}

During the CHB phase, we model the stellar wind of the VMS using the formulae from V01, whilst correcting for the proximity to the Eddington limit by fitting on the data from table 1 of V11. In this way we obtain a coefficient that allows us to convert the V01 rate to the $\Gamma\sscript{e}$ enhanced rates of stars approaching the Eddington limit \citep[similarly to][]{Chen15}. V11 modelled stars up to $300 M\sscript{\odot}$, however, as the logarithmic difference between the V11 and V01 rates shows little dependence on mass, we extrapolate this approach to higher masses. V11 state that their predicted wind velocities are a factor 2--4 less than derived empirically. The effect of rotation is also neglected. It should be noted that due to these two effects, and our extrapolation of the V11 models, we likely underestimate the mass loss of our VMSs. Therefore the masses of our VMSs and their resulting remnants should be taken as a conservative upper limit at solar metallicity.

During CHeB VMSs are depleted of hydrogen and are expected to show Wolf-Rayet like features. We follow the approach of \citet{Belkus07} and extrapolate the mass loss formula of \citet{Nugis00}:
\begin{equation}
	\log(\dot{M}) = -11  + 1.29 \log(L) +1.7 \log(Y) + 0.5 \log(Z).
\end{equation}

B07 explored models with Wolf-Rayet like mass loss rates (arbitrarily) up to 4 times weaker, which only left a remnant twice as massive. The uncertainty arising from extrapolation of this formula should be of little significance to the transport of young stars to the central parsec, as post main sequence VMSs are not massive enough to experience substantial dynamical friction after the cluster is disrupted (B07). However, if sufficiently chemically rejuvenated, a CHB VMS may be capable of bringing stars to the central parsec before losing most of its mass. Thus the evolution during the CHB stage is of most interest.

We make sure that in both burning phases the predicted mass loss never exceeds the photon tiring limit, the maximum mass loss rate that can theoretically be achieved using $100\%$ of the stars luminosity to drive the wind \citep{Owocki04}:
\begin{equation}
	\dot{M}\sscript{tir} = 0.032 \left(\frac{L}{10^6L\sscript{\odot}}\right) \left(\frac{R}{R\sscript{\odot}}\right) \left(\frac{M}{M\sscript{\odot}}\right)^{-1}.
\end{equation}
Here, the radii, $R$, of stars are taken from the mass-radius relation of \citet{Yungelson08}, which is in excellent agreement with \citet{Nadyozhin05}'s similarity theory models of VMSs, but requires less computational resources to calculate. The OB disk population is less than $7\unit{Myr}$ old. Hence, we assume approximately solar abundances such that $X\sscript{0}=0.7$, $Y\sscript{0}=0.28$ and $Z\sscript{0}=0.2$ \citep{Pols98}.

\subsubsection{Rejuvenation following collisions}

\citet{Nadyozhin05} show that VMSs have nearly all of their mass in their large convective cores. Repeated collisions can efficiently mix the core and the halo, keeping the star relatively homogeneous. The wind of the VMS also ensures homogeneity, as the loose radiative envelope is shedded by the stellar wind, leaving the surface with composition similar to the core.

We chemically rejuvenate a VMS following a collision with another star. We assume that stars colliding with the VMS efficiently mix with the convective core such that:

\begin{equation}
	X\sscript{new} = \frac{X\sscript{star}M\sscript{star} + X\sscript{VMS}M\sscript{VMS}}{M\sscript{VMS+star}}
	\label{Xmix.eq}
\end{equation}
Similarly for Y and Z. We approximate $X\sscript{star}(t)$ and $Y\sscript{star}(t)$ for main sequence stars by interpolating the detailed stellar models of \citet{Schaller92}. If a CHeB VMS collides with a hydrogen rich main sequence star, we assume that CHB is reignited. When two VMSs collide their composition is also assumed to be well mixed.

\section{Numerical Method}

\label{Numerical_Method.ch}
To model the effects of dynamical friction on self-consistent star cluster models we use the GPU-enabled direct $N$-body code NBODY6df \citep{Petts15}, which is a modified version of Aarseth's direct $N$-body code NBODY6 \citep{Aarseth1999,Nitadori12}. In this paper we model the background as an analytic stellar distribution with a central black hole (see section \S \ref{IC.ch}). In \citet{Petts15} we only tested our dynamical friction model for cases without a central black hole, however we discuss how to trivially add a black hole to the model in Appendix A. A validation of this approach via comparison with full $N$-body models computed with GADGET \citep{Springel01} is also given in the appendix.

We introduced an additional modification to the code to properly model the evolution of a VMS, as described in section \S \ref{VMS_evo.ch}. When a physical collision creates a star greater than $100\unit{M\sscript{\odot}}$ we flag it as a VMS and treat its evolution separately from the standard SSE package in NBODY6 \citep{Hurley00} via the method described in section \S \ref{VMS_evo.ch}. As the mass loss can be very large for VMSs, fine time resolution is needed to prevent overestimation of the mass loss. We introduce a new routine which integrates the mass and composition of the star between the dynamical time steps using a time step of $0.1 \unit{years}$, sufficiently accurate to resolve the evolution. V11 predict terminal wind speeds of a few thousand $\unit{km} \unit{s^{-1}}$ for VMSs, as such we assume that the stellar wind escapes the cluster and simply remove this mass from the VMS. An arbitrary number of VMSs can potentially form and evolve simultaneously in the simulation.

\section{Initial Conditions}
\label{IC.ch}

\begin{table}
\begin{minipage}{85mm}
\centering
  \begin{tabular}{@{}lrrrrr@{}}
   Name &$M\sscript{cl}$ &$r\sscript{hm}$&$W\sscript{0}$&$N$&$m\sscript{low}$\\
   &($10^5\unit{M_{\odot}}$)&($\unit{pc}$)&&&($\unit{M_{\odot}}$)\\
 \hline\hline
  1lo & 1.06 & 0.200 &6&32k&1.0\\ 
  1hi & 1.06 & 0.200 &6&128k&0.16\\ 
  1kr & 1.06 & 0.200 &6&186k&0.08\\

\hline

\end{tabular}
\caption{Initial conditions of the isolated simulations. Column 1 lists the name of the simulation. The naming convention is described in section \S \ref{IC.ch}. Columns 2,3 show the mass and half mass radius of the cluster. Column 4 shows the dimensionless central potential of the King model. Column 5 shows the number of particles, and column 6 gives the lower mass limit of the IMF. The upper mass limit is $100 \unit{M\sscript{\odot}}$ for all models.}
\label{iso.tbl}
\end{minipage}

\end{table}

\begin{table*}
\begin{minipage}{170mm}

\centering

  \begin{tabular}{@{}lrrrrrrrrr@{}}
   Name &$M\sscript{cl}$ &$r\sscript{hm}$&$W\sscript{0}$&$\bar{\rho}\sscript{c}$&$N$&$m\sscript{low}$& $f\sscript{bin}$&$R\sscript{a}$ & $v$\\
   &($10^5\unit{M_{\odot}}$)&($\unit{pc}$)&&($\unit{M_{\odot}pc^{-3}}$)&&($\unit{M_{\odot}}$)& ($\%$)&($\unit{pc}$)&($v\sscript{c}$) \\
 \hline\hline
    4lo15\_W4& 4.24  & 0.589 &4&$6.78\times10^5$& $2^{17}$&1.0&0 &15 & 1.0\\
  4lo15\_W4v75& 4.24  & 0.360 &4&$2.86\times10^6$& $2^{17}$&1.0&0 & 15 & 0.75\\  
  2lo10 & 2.12  & 0.220 &6&$1.32\times10^7$& $2^{16}$&1.0&0 & 10 & 1.0\\
  2lo10\_W4 & 2.12  & 0.318 &4&$2.09\times10^6$& $2^{16}$&1.0&0 & 10 & 1.0\\     
  2lo10\_v75 & 2.12  & 0.141 &6&$5.35\times10^7$& $2^{16}$&1.0&0 & 10 & 0.75\\  
  2lo10\_v5* & 2.12 & 0.220 &6&$1.32\times10^7$& $2^{16}$&1.0 &0 & 10 & 0.5\\
  2lo10\_v2* & 2.12  &0.220 &6&$1.32\times10^7$& $2^{16}$&1.0&0 & 10 & 0.2\\
  2lo5& 2.12 &0.135&6&$5.75\times10^7$& $2^{16}$&1.0&0 &  5 & 1.0\\
  2lu5& 2.12 &0.135&6&$5.75\times10^7$& 29k&1.0&0 &  5 & 1.0\\              
  1lo10 & 1.06  & 0.175 &6&$1.26\times10^7$& $2^{15}$&1.0&0 & 10 & 1.0\\
  1hi10 & 1.06  & 0.175 &6&$1.26\times10^7$& $2^{17}$&0.16&0 & 10 & 1.0\\
  1hi10\_b & 1.06  & 0.175 &6&$1.26\times10^7$& $2^{17}$&0.16&5 & 10 & 1.0\\  
  1lo10\_W4 & 1.06  & 0.273 &4&$1.69\times10^6$& $2^{15}$&1.0&0 & 10 & 1.0\\
  1lo10\_v75 & 1.06 & 0.112 &6&$4.94\times10^7$& $2^{15}$&1.0 &0 & 10 & 0.75\\
  1lo10\_v5* & 1.06  & 0.175 &6&$1.26\times10^7$& $2^{15}$&1.0&0 & 10 & 0.5\\
  1hi10\_v5* & 1.06  & 0.175 &6&$1.26\times10^7$& $2^{17}$&0.16&0 & 10 & 0.5\\  
  1lo10\_v2* & 1.06  & 0.175 &6&$1.26\times10^7$& $2^{15}$&1.0&0 & 10 & 0.2\\
  1hi10\_v2* & 1.06  & 0.175 &6&$1.26\times10^7$& $2^{17}$&0.16&0 & 10 & 0.2\\ 
  1hi10\_v2b* & 1.06  & 0.175 &6&$1.26\times10^7$& $2^{17}$&0.16&5 & 10 & 0.2\\
  1lo5 & 1.06  & 0.107 &6&$6.42\times10^7$& $2^{15}$&1.0&0 & 5 & 1.0\\
  1hi5 & 1.06  & 0.107 &6&$6.42\times10^7$& $2^{17}$&0.16&0 & 5 & 1.0\\
  1hi5\_b & 1.06  & 0.107 &6&$6.42\times10^7$& $2^{17}$&0.16&5 & 5 & 1.0\\        
  1hi5\_ms & 1.06  & 0.107 &6&$6.42\times10^7$& $2^{17}$&0.16&0 & 5 & 1.0\\
  1hi5\_W4d & 1.06  & 0.055 &4&$2.03\times10^8$& $2^{17}$&0.16&0 & 5 & 1.0\\   
  1lu5 & 1.06  & 0.107 &6&$6.42\times10^7$& 14.7k&1.0&0 & 5 & 1.0\\
 \hline

\end{tabular}
\caption{Initial conditions of the simulations. Column 1 lists the name of the simulation. The naming convention is described in section \S \ref{IC.ch}. Columns 2 and 3 give the mass and half mass radius of the cluster. Column 4 gives the dimensionless central potential of the King model. Column 5 shows the average core density. Column 6 shows the number of particles, and column 7 the lower mass limit of the IMF. The upper mass limit is $100 \unit{M\sscript{\odot}}$ for all models. Column 8 shows the initial binary fraction. Columns 9 and 10 show the initial position and velocity, in units of $\unit{pc}$ and the circular velocity, respectively. Models are set up to be Roche-filling at first pericentre passage; unless they are marked by an asterisk, in which case they are Roche-filling at their initial positions.}
\label{IC.tbl}

\end{minipage}
\end{table*}

In NBODY6df the background potential is assumed static and analytic; an assumption valid over the short timescales considered here (less than $7\unit{Myr}$). We adopt a Dehnen model \citep{Dehnen93}, representing the central region of the Galaxy. We use a slope $\gamma = 1.5$, scale radius $a = 8.625\unit{pc}$ and total mass $M\sscript{g} = 5.9\times 10^7 \unit{M_\odot}$, which closely reproduces the observed broken power-law profile obtained by \citet{Genzel03} for the central region of the Galaxy, yet has simple analytic properties. We place a central fixed point mass of $4.3\times10^6M\sscript{\odot}$ to represent SgrA* \citep{Gillessen09}.

\subsection{Physical and numerical constraints on the initial conditions}

\begin{figure}
 \includegraphics[width=\linewidth]{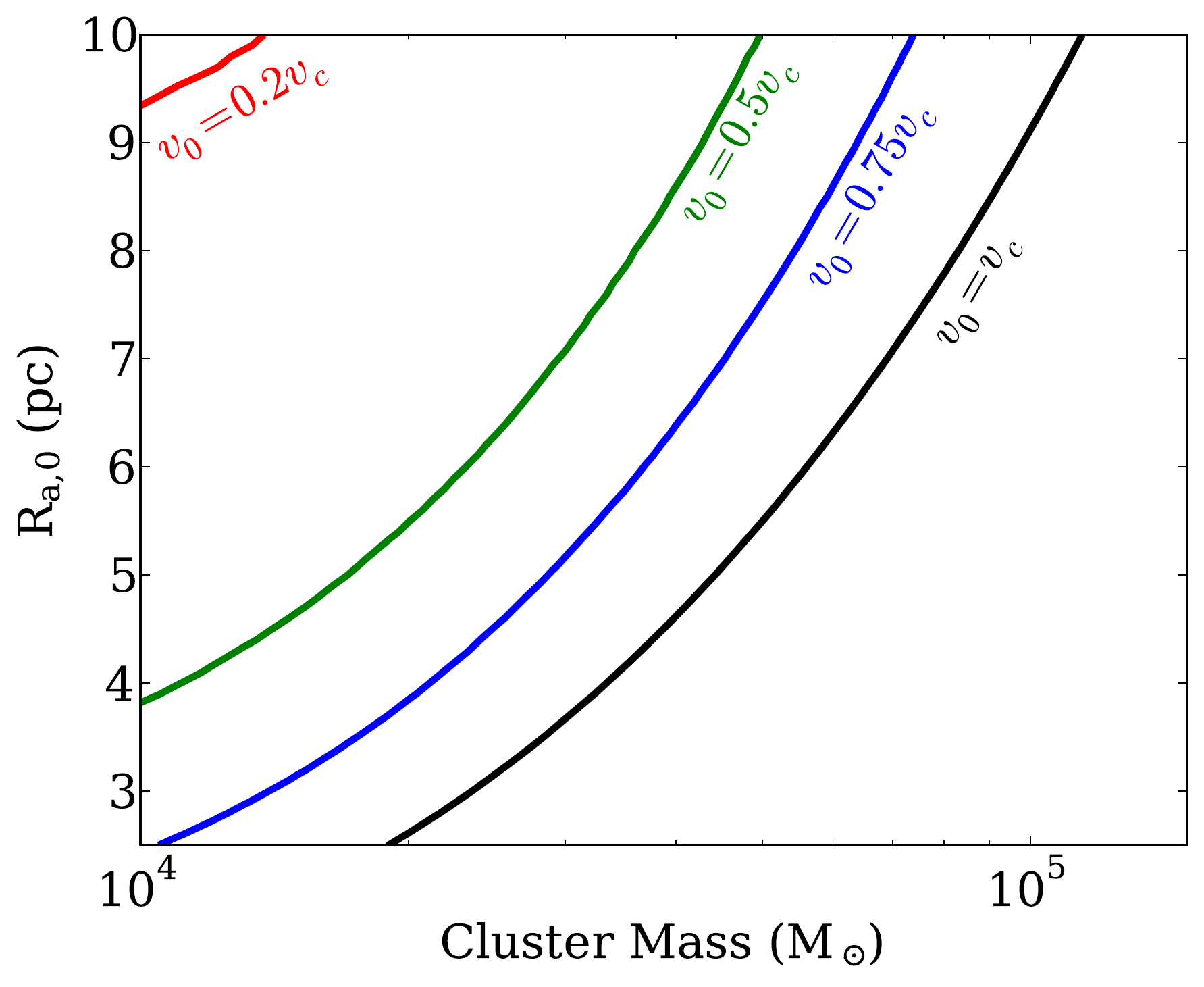}
 \caption{Contours of $T\sscript{df} = 7 \unit{Myr}$ as a function of cluster mass, initial distance, $R\sscript{a,0}$, and initial velocity, $v\sscript{0}$, given in units of the local circular velocity, $v\sscript{c}$. Models to the right of each line approach within $0.5\unit{pc}$ of SgrA* in less than $7\unit{Myr}$. The half mass radius of the cluster is assumed to be $0.1\unit{pc}$.} 
 \label{T_df.fig}
\end{figure}

There are two constraints on the initial conditions of the clusters. Firstly, they must reach the Galactic Centre within the age of the young populations. We therefore wish to model clusters that can potentially reach the Galactic Centre in less than $7\unit{Myr}$, so that we may test the migration model for both the clockwise disk and the S-stars. We obtain tight constraints on the initial orbital parameters by integrating the orbits of point masses in the Galactic Centre potential including dynamical friction. Fig. \ref{T_df.fig} shows contours of equal inspiral time for different initial masses, apocentres and initial velocities. Initial conditions to the right of each line are such that the clusters can reach within $0.5\unit{pc}$ in less than $7\unit{Myr}$. Arches like clusters \citep[initial mass $4-6\times10^4M\sscript{\odot}$,][]{Harfst10}) could reach the Galactic centre in less than $7\unit{Myr}$ if they formed at $\sim5\unit{pc}$, or from $7-10\unit{pc}$ if large initial eccentricities were assumed. More massive clusters can easily migrate $\sim 10\unit{pc}$ in $7\unit{Myr}$. We note that these inspiral times are lower limits, as real clusters would lose mass from stellar winds and tides. We choose to model only those clusters for which a point mass object of the same mass can reach the Galactic Centre within $\sim7\unit{Myr}$.

Secondly, the size of the clusters is limited by their small tidal limits when so close to SgrA*. Approximating the cluster as a point mass, the tidal radius is given by \citep{BT}:
\begin{equation}
	r^3\sscript{t} = \frac{GM\sscript{cl}}{\omega^2\sscript{p} +\left(\frac{d^2\Phi}{dR^2}\right)\sscript{p}},
\end{equation}
where $\omega\sscript{p}$ and $\left(\frac{d^2\Phi}{dR^2}\right)\sscript{p}$ are the angular velocity of a circular orbit and the second derivative of the potential at pericentre, respectively. The high mass requirement for fast inspiral, coupled with the small tidal limits, means that all models are inherently very dense and runaway mergers are expected. Although it is unknown whether such dense clusters are likely to form in the Galactic Centre, we explore these initial conditions in order to test the feasibility of the inspiral model.

\subsection{Initial Mass function}

\begin{figure}
 \includegraphics[width=\linewidth]{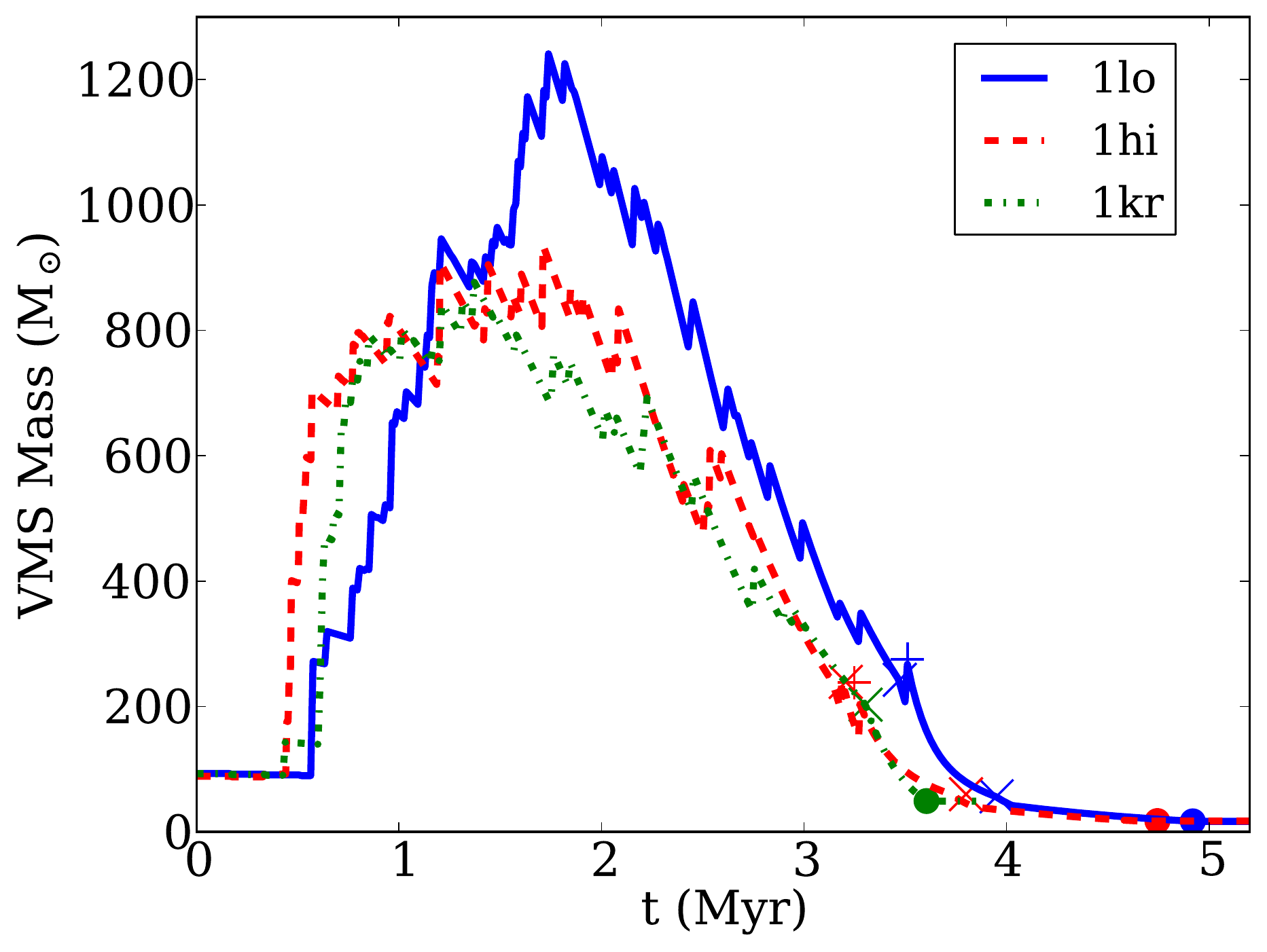}
 \caption{Time evolution of the VMS mass formed in simulations with different lower mass cutoffs in the IMF. The solid blue, dashed red and dot-dashed green lines show the mass of the VMS in simulations ${\tt 1lo}$, ${\tt 1hi}$ and ${\tt 1kr}$, respectively. The diagonal crosses show the end of CHB, the vertical crosses show re-ignition of CHB, and the solid circles show where the remnants collapse to black holes.} 
 \label{mf_test.fig}
\end{figure}

We sample stars from a Kroupa initial mass function (IMF) with an upper limit of $100 \unit{M\sscript{\odot}}$ \citep{Kroupa01}. A lower mass limit of $0.08 \unit{M\sscript{\odot}}$ would yield the most physically realistic results, but at a computational cost unfeasible for a parameter study of such massive clusters at the current time  (365k-730k particles for the most massive models explored). However, truncating the low end of the IMF means that one samples too many massive stars as compared with a full Kroupa IMF. To quantify the difference this has on VMS formation, we ran three test simulations at different mass resolutions, in the absence of a tidal field. Simulations ${\tt 1lo}$, ${\tt 1hi}$ and ${\tt 1kr}$ have lower cutoffs of $1.0$, $0.16$ and $0.08 \unit{M\sscript{\odot}}$, respectively. We model star clusters as King models with dimensionless central potential, $W\sscript{0} = 6$, and with no primordial mass segregation. The parameters of the isolated simulations are displayed in Table \ref{iso.tbl}. Fig. \ref{mf_test.fig} shows the VMS mass as a function of time for simulations ${\tt 1lo}$, ${\tt 1hi}$ and ${\tt 1kr}$, showing that better sampling of the low end of the IMF inhibits the growth of the VMS. This occurs because primarily high mass stars build up the VMS, due to their short dynamical friction timescales and large cross sections for collision. In simulation ${\tt 1kr}$, although half the cluster mass is comprised of stars less massive than $0.58 \unit{M\sscript{\odot}}$, only 37 stars less massive than $0.58 \unit{M\sscript{\odot}}$ are consumed throughout the entire lifetime of the VMS. The VMS initially grows very rapidly. However, the late main sequence evolution is dominated by the strong stellar wind of the helium rich VMS. Throughout its lifetime, the VMS in simulation ${\tt 1kr}$ removes $2244\unit{M\sscript{\odot}}$ of material from the cluster through its stellar wind, $\sim 2\%$ of the cluster mass. During CHeB, simulations ${\tt 1lo}$ and ${\tt 1hi}$ reignite CHB via collision with a massive main sequence star, resulting in a lower remnant mass at collapse. The late evolution is very stochastic, however this is not important for the migration of young stars to the Galactic Centre, as the VMS only provides gravitational binding energy comparable to normal cluster stars during its CHeB phase.

Fig. \ref{mf_test.fig} shows that a lower limit of $0.16 \unit{M\sscript{\odot}}$ is sufficient to resolve the mass evolution of the VMS, and as we are only interested in the final distribution of OB stars, this IMF is sufficient for our simulations. We cannot evolve the most massive clusters at high mass resolution, as these models become too computationally expensive. As a compromise, we test a large range of initial conditions with a lower limit of $1 \unit{M\sscript{\odot}}$, and re-run a selection of initial conditions with a lower limit of $0.16 \unit{M\sscript{\odot}}$ to obtain more realistic results. We can simultaneously use the low resolution simulations to explore the possibility of an initially top heavy mass function for clusters forming close to SgrA*. A very top heavy function is observed for the clockwise disk \citep{Lu13}, however, it is unknown whether a top heavy IMF is expected from the collapse of GMCs at $\sim5-10 \unit{pc}$ from SgrA*.

\subsection{Binary fraction}

\label{binarity.ch}
Some simulations include a population of primordial binaries. Binaries are initialised as follows. Firstly all stars more massive than $5 \unit{M\sscript{\odot}}$ are ordered by mass. The most massive star is then paired with the second most massive star, and so on. This choice is motivated by observational data showing that massive OB stars are more likely to form in binary systems with mass-ratios of order unity \citep{Kobulnicky07,Sana11}. Once all stars more massive than $5 \unit{M\sscript{\odot}}$ are in binaries, lower mass stars are paired at random until the specified binary fraction (the fraction of stars initially in a binary system) is reached \citep{Kroupa08}. For stars more massive than $5 \unit{M\sscript{\odot}}$, the periods and eccentricities are drawn from the empirical distributions derived in \citet{Sana11}, which show that short periods and low eccentricities are preferred in massive binaries. For lower mass stars the periods are drawn from the \citet{Kroupa95} period distribution and are assigned thermal eccentricities. The mass of a binary and its initial position in the cluster are assumed to be independent.

\subsection{Simulations}
The initial conditions are described in Table \ref{IC.tbl} and are referred to by the following naming convention: ${\tt <M><mf><R\sscript{a}>}$, where $<\rm{M}>$ is the cluster mass in units of $\sim 10^5 \unit{M\sscript{\odot}}$, $ <\rm{mf}>$ is the mass resolution of the simulation, and $<\rm{R\sscript{a}}>$ is the initial galactocentric distance in pc. For most simulations  we sample from a Kroupa IMF with an upper mass cut off, $m\sscript{up} = 100\unit{M\sscript{\odot}}$. The ``$\rm{\textit{lo}}$'' resolution models have a lower mass cut off, $m\sscript{low} = 1\unit{M\sscript{\odot}}$ and mean mass, $\left\langle\ m\sscript{*}\right\rangle= 3.26\unit{M\sscript{\odot}}$. The ``$\rm{\textit{hi}}$'' resolution models have $m\sscript{low} = 0.16\unit{M\sscript{\odot}}$ and $\left\langle\ m\sscript{*}\right\rangle= 0.81\unit{M\sscript{\odot}}$. For simulations with $ <\rm{mf}> = \rm{\textit{lu}}$ we use an IMF identical to the mass function of the clockwise disk \citep{Lu13}. The simulation name is followed by a suffix describing additional information about the simulation. The suffix $\rm{W4}$ denotes that the dimensionless central potential, $W\sscript{0}$, is initially 4 instead of 6. The suffix $\rm{vX}$ indicates an eccentric orbit with initial velocity, $\rm{0.X} v\sscript{c}$ (where $v\sscript{c}$ is the circular velocity at the initial position). The Suffix ``ms'' indicates that the cluster is primordially mass segregated. Finally, the suffix ``b'' denotes the inclusion of primordial binaries (see section \S \ref{binarity.ch}). Most models are Roche-filling at first pericentre passage, apart from runs marked with an asterisk, which are Roche-filling at their initial positions. The model with the suffix ``d'' is extremely Roche under-filling at its initial position.

\section{Results}
\label{results.ch}

\begin{table}
\begin{minipage}{85mm}
\centering
  \begin{tabular}{@{}lrrrrr@{}}
   Run Name &$M\sscript{VMS,max}$&$M\sscript{VMS,rem}$&$t\sscript{VMS}$\\
   &$\unit{M\sscript{\odot}}$&$\unit{M\sscript{\odot}}$& $\unit{Myr}$\\
 \hline
 4lo15\_W4 &-&-&-\\
 4lo15\_W4v75 &139.1&19.8&6.16\\
 2lo10 &3472.9&85.7&3.26\\
 2lo10\_W4 &703.2&53.0&4.86\\
 2lo10\_v75 &3858.4&79.5&2.98\\
 2lo10\_v5* &2069.5&119.7&2.92\\
 2lo10\_v2* &-&-&-\\
 2lo5 &2804.1&169.0&2.78 \\
 2lu5 &4086.6&378.5&2.54\\
 1lo10 &1541.8&73.5&3.11\\
 1hi10 &886.903&48.7&3.77\\
 1hi10\_b &1232.4&17.3&3.94 \\
 1lo10\_W4 &721.1&20.6&5.05\\
 1lo10\_v75 &3413.8&199.8&2.84\\
 1lo10\_v5* &1462.2&82.2&2.77\\
 1hi10\_v5* &927.5&56.8&2.92\\
 1lo10\_v2* &125.3&16.4&4.37\\
 1hi10\_v2* &234.9&22.7&3.07\\
 1hi10\_v2b* &481.3&32.6&2.58\\
 1lo5 &3310.7&245.6&2.60\\
 1hi5 &1415.7&62.0&2.44\\
 1hi5\_b &1747.1&80.1&2.52\\
 1hi5\_ms &1338.8&65.8&2.99\\
 1hi5\_W4d &2044.4& 77.8&2.91\\
 1lu5 &3950.6&355.1&2.42\\

\hline

\end{tabular}
\caption{Properties of VMSs formed in the simulations. Column 1 lists the name of the simulation. Column 2 shows the maximum mass via stellar collisions, column 3 shows  the resulting remnant mass after CHeB, and column 4 shows the epoch in the cluster evolution when the VMS collapses.}
\label{VMS.tbl}
\end{minipage}

\end{table}

In all models, the clusters are completely tidally disrupted in less than $7\unit{Myr}$. Massive clusters migrate farther in than lower mass clusters on the same initial orbits, due to shorter dynamical friction timescales and less efficient tidal stripping. However any cluster that reaches $\sim 3 \unit{pc}$ is rapidly dissolved by its shrinking tidal limit as it approaches SgrA*. Clusters on eccentric orbits inspiral faster, as they pass through denser regions of the cusp periodically. However, clusters on very eccentric orbits (e.g. {$\tt 2lo10\_v2*$}, $e \sim 0.9$) disrupt on the first few pericentre passages, depositing stars at large distances along the initial cluster trajectory.

Most simulations naturally form a VMS in less than $1\unit{Myr}$ due to their high initial densities. However, the initial rapid mass accretion soon loses to the increasing mass loss rate and relaxation of the cluster, causing the VMS to collapse to a black hole of $\sim 20 - 250 \unit{M\sscript{\odot}}$ after $2-5 \unit{Myr}$ ($300-400 \unit{M\sscript{\odot}}$ for models with a \citet{Lu13} IMF), typically before their parent clusters completely disrupt. Table \ref{VMS.tbl} shows the maximum mass, remnant mass and lifetime of the VMS formed in each simulation. The clusters become completely unbound at $\sim 2-3 \unit{pc}$, and the IMBHs formed are not massive enough to experience significant dynamical friction and drag stars close to SgrA* (dynamical friction timescales for even a $400 \unit{M\sscript{\odot}}$ IMBH are longer than $100 \unit{Myr}$). Conversely, the evolution of the VMS does not appear to significantly inhibit the inspiral of the cluster, as only $\lesssim 2\%$ of the initial cluster mass is typically expelled by the VMS throughout its lifetime.

\begin{table*}
\begin{minipage}{170mm}
\centering
  \begin{tabular}{@{}lrrrrrrrrr@{}}
   Run Name & $\left\langle a \right\rangle\sscript{all}$ & $\left\langle a \right\rangle\sscript{>8\unit{M\sscript{\odot}}}$ &$N(<1\unit{pc})$&$N(<1\unit{pc},>8\unit{M\sscript{\odot}})$&$\left\langle D\sscript{2D} \right\rangle\sscript{7 \unit{Myr}}$&$\left\langle D\sscript{2D} \right\rangle\sscript{15 \unit{Myr}}$\\
      &(pc)&(pc)&&&(pc)&(pc)\\
 \hline 
 4lo15\_W4 &9.20 $\pm$ 4.15& 8.86 $\pm$ 3.94& 0& 0& 9.22 $\pm$ 4.90& 9.31 $\pm$ 4.95\\
 4lo15\_W4v75 &5.23 $\pm$ 3.27& 4.89 $\pm$ 3.11& 0& 0& 5.17 $\pm$ 3.88& 5.34 $\pm$ 4.00\\
 2lo10 &6.90 $\pm$ 2.92& 6.19 $\pm$ 2.76& 0& 0& 6.24 $\pm$ 3.00& 6.38 $\pm$ 3.03\\
 2lo10\_W4 &6.43 $\pm$ 2.70& 6.00 $\pm$ 2.62& 0& 0& 6.05 $\pm$ 2.86& 6.13 $\pm$ 2.85\\
 2lo10\_v75 &4.61 $\pm$ 2.30& 4.20 $\pm$ 2.09& 0& 0& 4.36 $\pm$ 2.56& 4.44 $\pm$ 2.55\\
 2lo10\_v5* &5.51 $\pm$ 2.86& 5.37 $\pm$ 2.83& 0& 0& 6.31 $\pm$ 4.50& 6.40 $\pm$ 4.62\\
 2lo10\_v2* &5.52 $\pm$ 3.17& 5.44 $\pm$ 3.14& 15& 0& 6.95 $\pm$ 5.48& 6.85 $\pm$ 5.44\\
 2lo5 &2.53 $\pm$ 1.33& 2.23 $\pm$ 1.23& 3970& 270& 2.26 $\pm$ 1.36& 2.32 $\pm$ 1.31\\
 2lu5 &3.33 $\pm$ 1.54& 3.24 $\pm$ 1.50& 3& 0& 3.29 $\pm$ 1.64& 3.32 $\pm$ 1.68\\
 1lo10 &8.79 $\pm$ 2.49& 8.24 $\pm$ 2.35& 0& 0& 8.29 $\pm$ 2.57& 8.47 $\pm$ 2.67\\
 1hi10 &8.02 $\pm$ 2.49& 7.18 $\pm$ 2.25& 0& 0& 7.22 $\pm$ 2.40& 7.47 $\pm$ 2.54\\
 1hi10\_b &8.76 $\pm$ 2.49& 8.71 $\pm$ 3.38& 0& 0& 8.07 $\pm$ 2.58& 8.08 $\pm$ 2.76\\
 1lo10\_W4 &8.88 $\pm$ 2.53& 8.40 $\pm$ 2.29& 0& 0& 8.43 $\pm$ 2.48& 8.49 $\pm$ 2.53\\
 1lo10\_v75 &6.42 $\pm$ 2.09& 6.19 $\pm$ 1.87& 0& 0& 6.42 $\pm$ 2.51& 6.54 $\pm$ 2.60\\
 1lo10\_v5* &6.09 $\pm$ 2.36& 5.90 $\pm$ 2.20& 0& 0& 6.93 $\pm$ 3.68& 6.94 $\pm$ 3.74\\
 1hi10\_v5* &5.87 $\pm$ 2.32& 5.77 $\pm$ 2.20& 0& 0& 6.67 $\pm$ 3.71& 6.66 $\pm$ 3.67\\
 1lo10\_v2* &5.64 $\pm$ 2.87& 5.56 $\pm$ 2.89& 9& 0& 7.56 $\pm$ 5.38& 7.44 $\pm$ 5.19\\
 1hi10\_v2* &5.62 $\pm$ 2.88& 5.58 $\pm$ 2.79& 10& 0& 7.63 $\pm$ 5.43& 7.82 $\pm$ 5.76\\
 1hi10\_v2b* &5.70 $\pm$ 2.90& 6.56 $\pm$ 3.38& 8& 1& 8.13 $\pm$ 5.51& 8.00 $\pm$ 5.35\\
 1lo5 &4.04 $\pm$ 1.36& 3.81 $\pm$ 1.28& 0& 0& 3.85 $\pm$ 1.51& 3.90 $\pm$ 1.55\\
 1hi5 &3.60 $\pm$ 1.33& 3.18 $\pm$ 1.25& 0& 0& 3.23 $\pm$ 1.39& 3.24 $\pm$ 1.31\\
 1hi5\_b &3.98 $\pm$ 1.36& 4.62 $\pm$ 2.69& 0& 0& 3.62 $\pm$ 1.24& 3.71 $\pm$ 1.27\\
 1hi5\_ms &4.24 $\pm$ 1.47& 3.20 $\pm$ 1.11& 0& 0& 3.25 $\pm$ 1.37& 3.30 $\pm$ 1.48\\
 1hi5\_W4d &3.08 $\pm$ 1.27& 2.56 $\pm$ 1.13& 0& 0& 2.63 $\pm$ 1.53& 2.64 $\pm$ 1.49\\
 1lu5 &4.36 $\pm$ 1.38& 4.27 $\pm$ 1.35& 0& 0& 4.30 $\pm$ 1.44& 4.34 $\pm$ 1.41\\

\hline

\end{tabular}
\caption{Final distributions of stars originally from the cluster. Simulations are run until the cluster is completely unbound and any VMSs collapse, up to a maximum of 7 Myr. Column 1 shows the name of each simulation. Column 2 shows the mean semi-major axis of all stars remaining in the simulation and the standard deviation. Column 3 shows the same for stars more massive than $8\unit{M\sscript{\odot}}$ that are still on the main sequence at $7 \unit{Myr}$. Column 4 shows the total number of  stars with final semi-major axes less than $1\unit{pc}$, and column 5 shows only those which are are still on the main sequence and more massive than $8\unit{M\sscript{\odot}}$ at $7 \unit{Myr}$. Columns 6 and 7 show the distributions of projected distances from SgrA* for main sequence stars visible at 7 and $15 \unit{Myr}$, respectively. The dissolved clusters are projected so that the resulting disk of stars rotates clockwise in the sky. }
\label{dist.tbl}
\end{minipage}

\end{table*}

For each model, Table \ref{dist.tbl} shows the final distribution of stars after complete cluster dissolution and death of any VMSs. We show the distributions of semi-major axes for all stars and main sequence stars more massive than $8\unit{M\sscript{\odot}}$ at $7 \unit{Myr}$, as well as how many of these stars have final semi-major axes smaller than $1 \unit{pc}$. We use a $8\unit{M\sscript{\odot}}$ cut-off as these are the faintest main sequence stars spectroscopically observable in the Galactic Centre with current telescopes, $K \geq 15.5$ \citep{Do09,Do13,Lu13,Feldmeier15}. Although photometric studies can see objects down to magnitudes of $K<19-18$ \citep[$\sim 2 \unit{M\sscript{\odot}}$ main sequence stars,][]{Genzel03}, it is impossible to determine whether these stars are young or old. We also show the projected distributions of visible main sequence stars at $7 \unit{Myr}$ and $15 \unit{Myr}$ (as the S-star population may be older than the disk population, see section \S \ref{Introduction.ch}).

\subsection{Low Resolution Models}

\label{results_lowres.ch}

\begin{figure}
 \includegraphics[width=\linewidth]{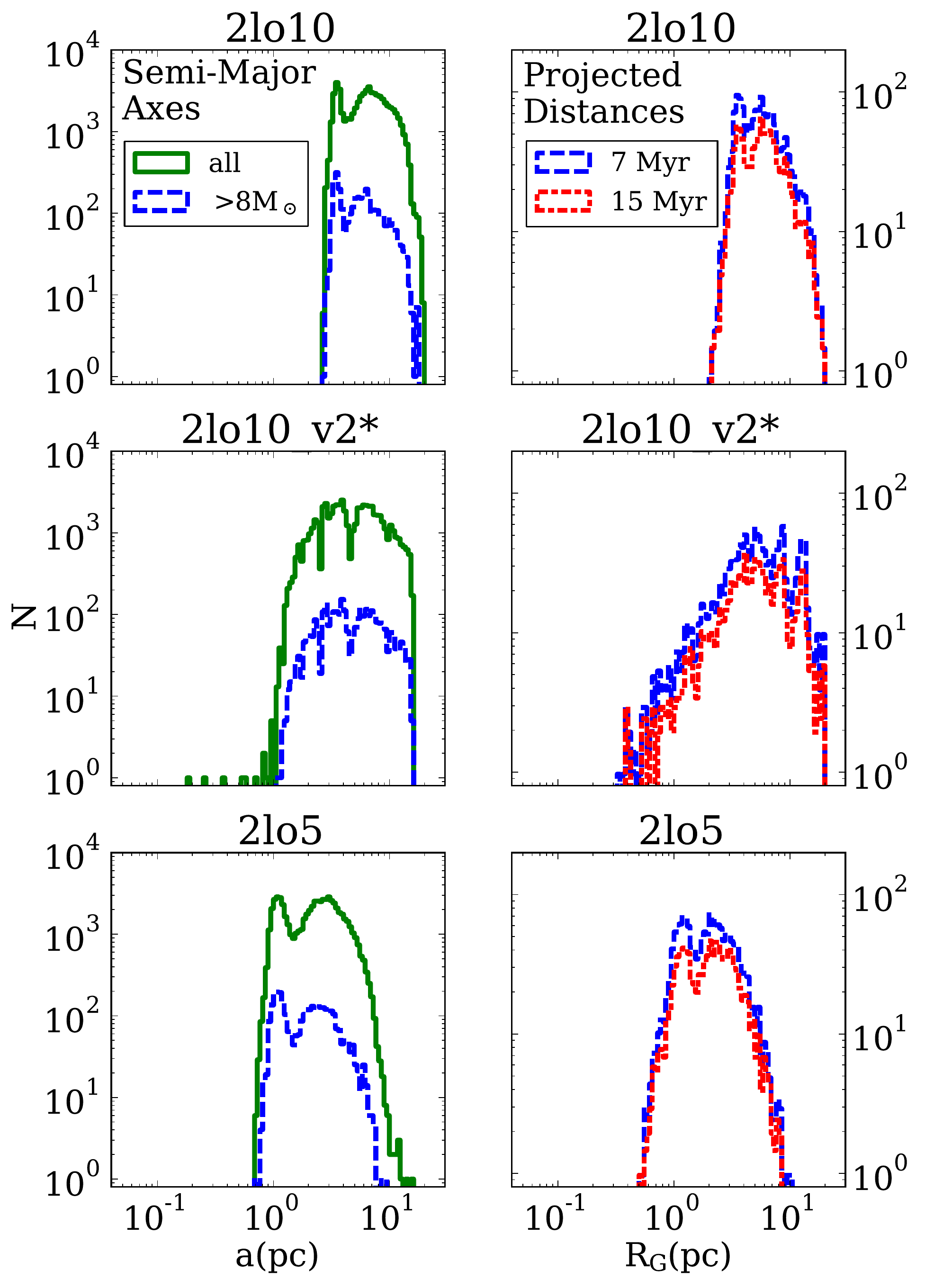}
 \caption{Left: Final distributions of semi-major axes of stars in simulations {$\tt 2lo10$}, {$\tt 2lo10\_v2*$} and {$\tt 2lo5$} at $T=7\unit{Myr}$. The solid green histogram shows all the stars. The dashed blue histogram shows main sequence stars more massive than $8\unit{M\sscript{\odot}}$ at $T=7 \unit{Myr}$. Right: Final distributions of the projected distances of stars from SgrA* in the same simulations. The dashed blue and dot-dashed red histograms show the distributions of main sequence stars more massive than $8\unit{M\sscript{\odot}}$ at $T=7\unit{Myr}$ and $T=15\unit{Myr}$, respectively, projected to rotate clockwise on the sky. The y-values of the projected distributions are re-normalised to the expected number of stars had the model been simulated with a full Kroupa IMF. } 
 \label{dist64k.fig}
\end{figure}

Fig. \ref{dist64k.fig} shows the final distributions of the semi-major axes and projected positions of stars for a representative selection of the models with a lower mass cutoff of $1 \unit{M\sscript{\odot}}$ ($<\rm{mf}> = \rm{lo}$). In all models the final distributions are broad, with a standard deviation of $\sim 2 \unit{pc}$. Other simulations show similar distributions, with less massive and less eccentric models dissolving farther out (see Table \ref{dist.tbl}).

Models with very eccentric orbits (e.g. {$\tt 2lo10\_v2*$}) can bring stars close to SgrA*, however, very few stars have final semi-major axes smaller than $1\unit{pc}$. No stars more massive than $8 \unit{M\sscript{\odot}}$ are scattered to semi-major axes smaller than $1 \unit{pc}$ in either {$\tt 1lo10\_v2*$} or {$\tt 2lo10\_v2*$}. This is likely due to the preferential loss of low mass stars, whereas high mass stars remain inside the cluster for longer, and end up tracing the final cluster orbit.

Simulation {$\tt 2lo5$} is the only non-radial model to bring stars to the central parsec, and the only model that brings a significant number of massive stars. However, one would expect to also see $\sim 3000$ massive stars in the range $1-10 \unit{pc}$, about $10$ times more than reach the central parsec. The right side of Fig. \ref{dist64k.fig} shows the distributions of projected distances of stars that are spectroscopically visible at $7$ and $15 \unit{Myr}$. The amplitudes of the distributions are normalised to the expected number of stars had the simulation been run with a Kroupa IMF. The stars are projected to rotate clockwise in the sky. It can be seen that for all simulations, more than $1000$ young stars are observed out to $\sim 10 \unit{pc}$. Considering current observational limitations, if a cluster were present in the central $\sim 10 \unit{pc}$ within the last $\sim 15 \unit{Myr}$, a large number of stars would be observable up to $\sim 10 \unit{pc}$, suggesting it is unlikely that any clusters have inhabited this region in the last $\sim 15 \unit{Myr}$.

Simulations {$\tt 2lo10$} and {$\tt 2lo10\_W4$} have the same initial orbit and mass, yet {$\tt 2lo10\_W4$} is less concentrated. The lower density and longer relaxation timescale cause {$\tt 2lo10\_W4$} to form a less massive VMS than {$\tt 2lo10$}. However, the VMS in {$\tt 2lo10\_W4$} lives longer as not all the most massive stars are consumed within $\sim 1 \unit{Myr}$. The models end up with similar final distributions of the resulting disk (see Table \ref{dist.tbl}). The same trend is seen for the less massive analogues, {$\tt 1lo10$} and {$\tt 1lo10\_W4$}. The two most massive simulations {$\tt 4lo10\_W4$} and {$\tt 4lo10\_W4v75$}, are massive enough to reach the central parsec from $15 \unit{pc}$ in $\sim 7 \unit{Myr}$, but with central densities low enough to suppress the formation of VMSs. However, these simulations are more susceptible to tides, and are tidally disrupted at large radii.

\subsection{Higher resolution models}

\begin{figure}
 \includegraphics[width=\linewidth]{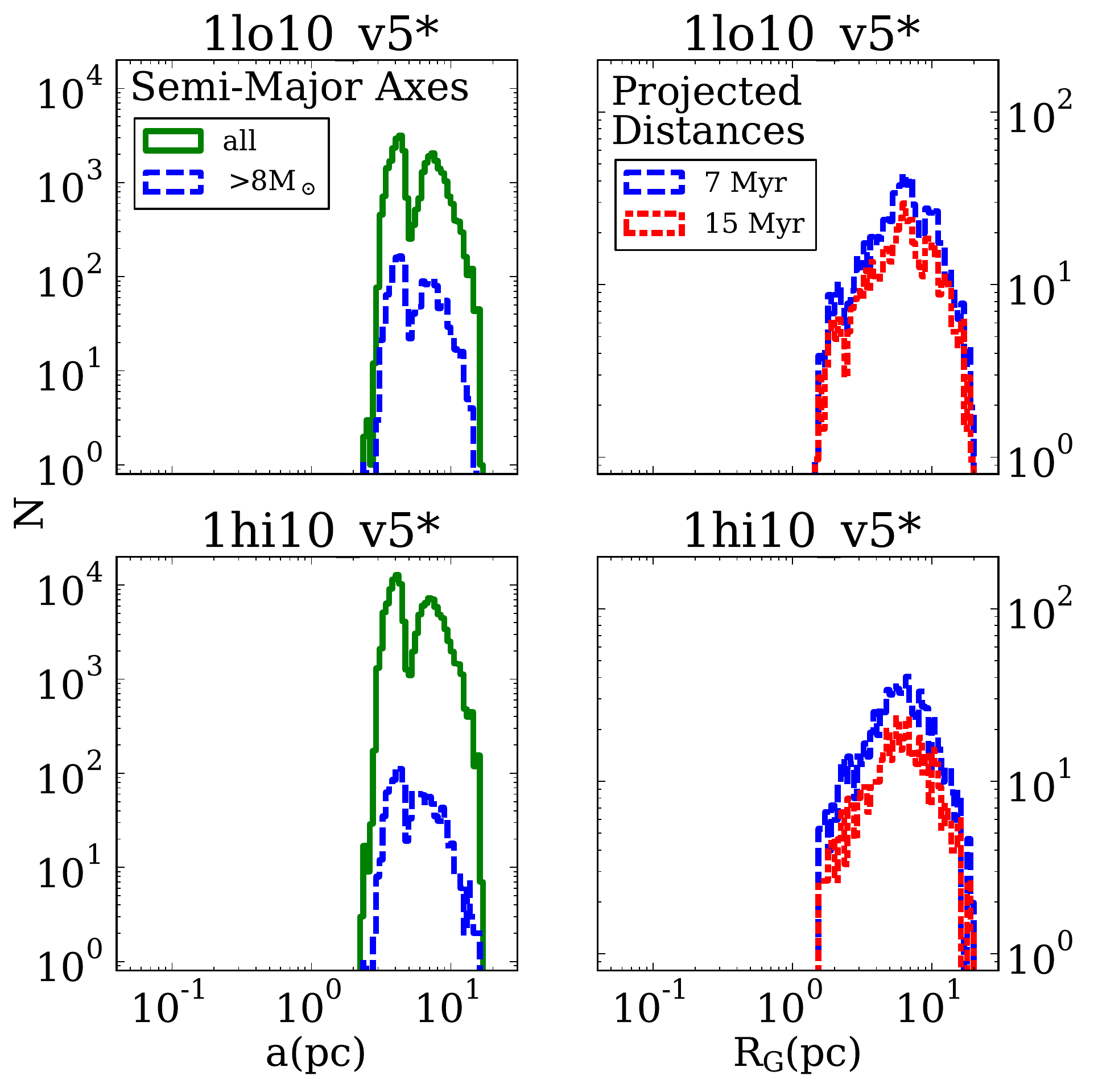}
 \caption{Comparison between simulations {$\tt 1lo10\_v5*$} and {$\tt 1hi10\_v5*$}, which have almost identical initial conditions, the latter sampling better the low end of the IMF. The left panels show the distribution of semi-major axes of all stars (solid green line) and stars more massive than $8\unit{M\sscript{\odot}}$ at $T=7\unit{Myr}$ (dashed blue line). The right panels show the distributions of projected distances of main sequence stars more massive than $8\unit{M\sscript{\odot}}$ at $T=7\unit{Myr}$ (dashed blue) and $T=15\unit{Myr}$ (dot-dashed red). The stars are projected so that the disk rotates clockwise in the sky. The y-values of the projected distributions are re-normalised to the expected number of stars had the model been simulated with a full Kroupa IMF.} 
 \label{1hi10_v5_s.fig}
\end{figure}

Fig. \ref{1hi10_v5_s.fig} shows a comparison between simulations {$\tt 1lo10\_v5*$} and {$\tt 1hi10\_v5*$}, which have the same initial conditions, except {$\tt 1hi10\_v5*$} better samples the low mass end of the IMF. The panels on the left show the distributions of semi-major axes for all the stars and main sequence stars more massive than $8\unit{M\sscript{\odot}}$ at $7 \unit{Myr}$. The distributions are very similar, however {$\tt 1hi10\_v5*$} has a smaller ratio of spectroscopically visible stars to all stars due to differences in the IMF. The panels on the right show the projected distributions of main sequence stars visible at $7$ and $15 \unit{Myr}$. For the projected distributions, the number of stars is re-normalised to the expected number of stars had the simulation been run with a Kroupa IMF from $0.08-100 \unit{M\sscript{\odot}}$. Although massive stars are consumed to construct the VMS, this is a small fraction of the population. The distributions look very similar in shape and magnitude, indicating that models run with a lower limit of $1 \unit{M\sscript{\odot}}$ produce similar final distributions to simulations that better sample the IMF. This verifies the validity of the normalisation approach used on the projected visible distributions in Fig. \ref{dist64k.fig}.

\begin{figure}
 \includegraphics[width=\linewidth]{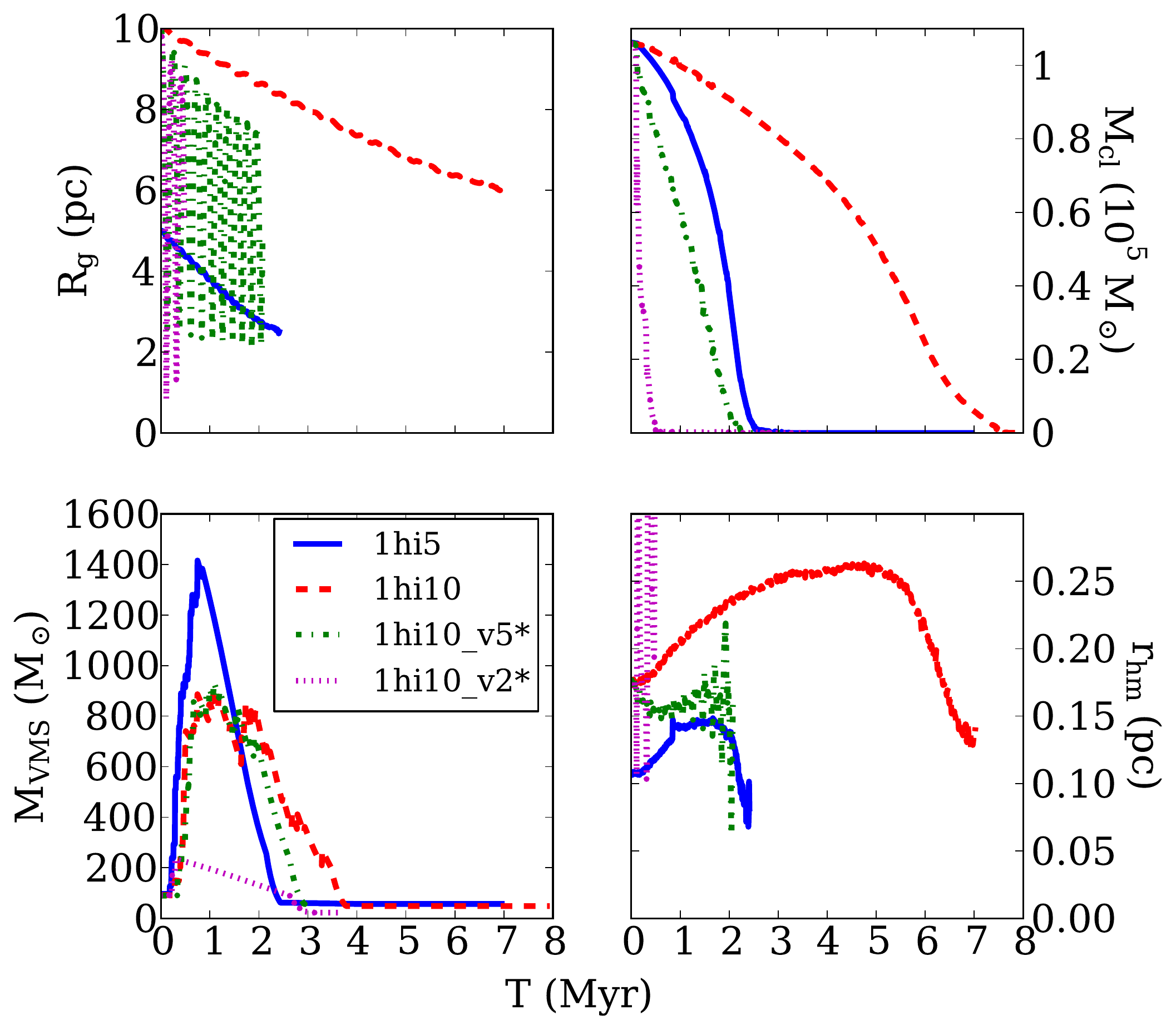}
 \caption{The evolution of galactocentric distance, cluster mass, VMS mass and cluster half mass radius as a function of time, for simulations {$\tt 1hi5$} (solid blue lines), {$\tt 1hi10$} (dashed red lines), {$\tt 1hi10\_v5*$} (dot-dashed green lines) and {$\tt 1hi10\_v2*$} (dotted magenta lines).} 
 \label{hievo.fig}
\end{figure}

Fig. \ref{hievo.fig} demonstrates how simulations {$\tt 1hi5$}, {$\tt 1hi10$}, {$\tt 1hi10\_v5*$} and {$\tt 1hi10\_v2*$} evolve with time. The top two panels show the evolution of the Galactocentric distance of the cluster and the mass enclosed within the tidal radius. The bottom two panels show the evolution of the VMS mass and the half mass radius of the cluster. Simulations {$\tt 1hi5$}, {$\tt 1hi10$} and {$\tt 1hi10\_v5*$} quickly form a VMS and expand due to rapid two body relaxation in the dense core. The expansion lowers the core density and thus the collision rate. The reduced collision rate allows the VMSs to rapidly burn their fuel and collapse without significant hydrogen rejuvenation. Simulation ${\tt 1hi5}$ forms a more massive VMS than the other simulations as it is initially $\sim 10$ times as dense, however the resulting increased luminosity decreases its lifetime. In simulation {$\tt 1hi10\_v2*$}, the cluster becomes unbound before the massive stars can reach the centre of the cluster, however the initial density is high enough that a $235 \unit{M\sscript{\odot}}$ VMS forms by the first pericentre passage. The self-limiting nature of the VMS formation is discussed in section \S \ref{discussion.ch}.

\subsection{Models with extreme initial conditions}

\begin{figure}
 \includegraphics[width=\linewidth]{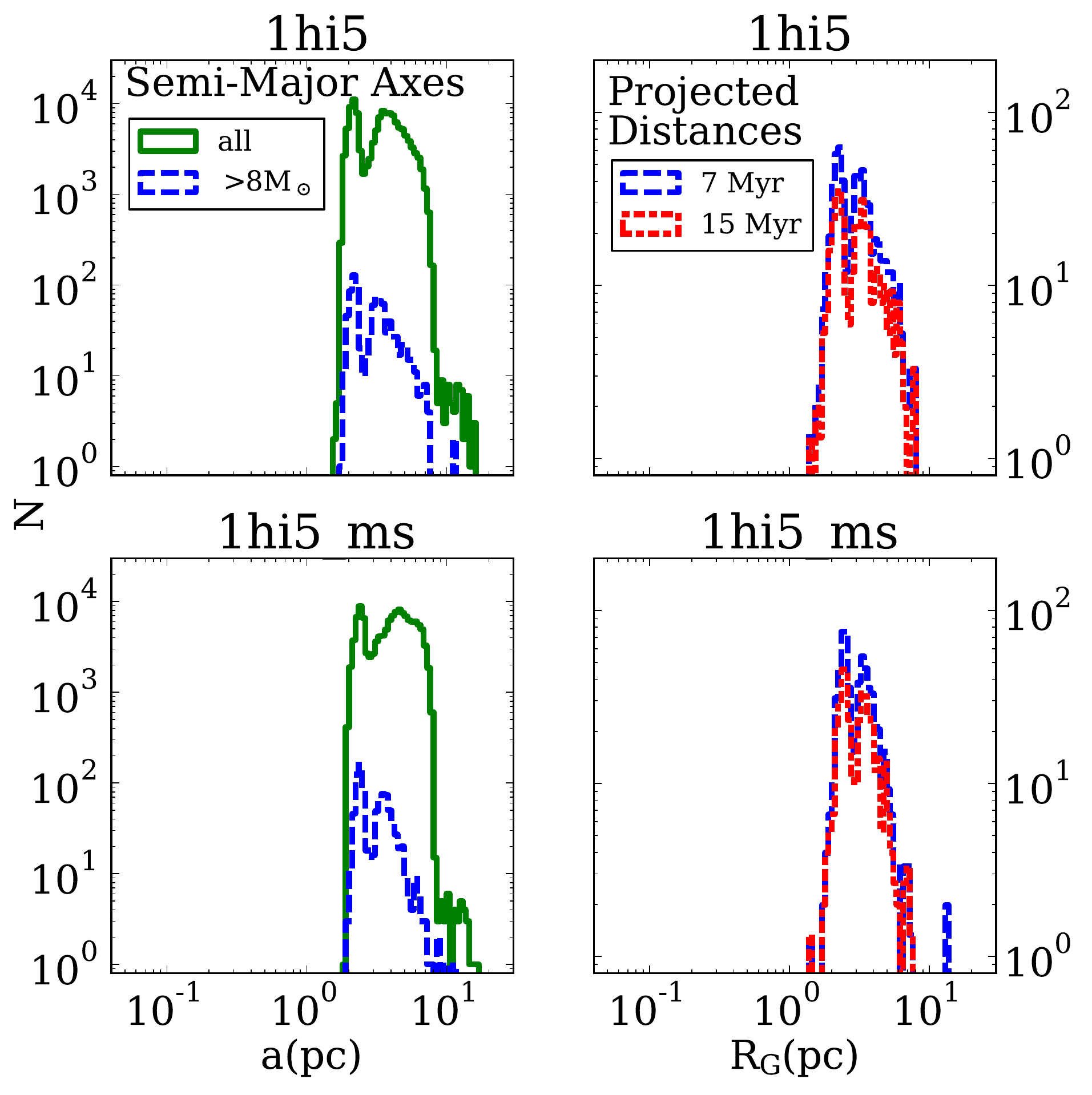}
 \caption{Comparison between simulations {$\tt 1hi5$} and {$\tt 1hi5\_ms$}, which have almost identical initial conditions, however the latter is primordially mass segregated. The left panels show the distribution of semi-major axes of all stars (solid green line) and stars more massive than $8\unit{M\sscript{\odot}}$ at $T=7\unit{Myr}$ (dashed blue line). The right panels show the distributions of projected distances of spectroscopically visible stars at $T=7\unit{Myr}$ (dashed blue) and $T=15\unit{Myr}$ (dot-dashed red). The stars are projected so that the disk rotates clockwise in the sky. The y-values of the projected distributions are re-normalised to the expected number of stars had the model been simulated with a full Kroupa IMF.} 
 \label{mass_seg.fig}
\end{figure}

The young clockwise disk population exhibits a top heavy mass function, with power law index $\alpha \sim 1.7$ \citep{Lu13}. In the context of the cluster inspiral scenario this has been explained by mass segregation inside the cluster, with the most massive stars reaching the central parsec, and low mass stars being preferentially lost due to tides during inspiral. However, as we have shown in section \S \ref{results_lowres.ch}, clusters lose massive stars as well as low mass stars throughout inspiral, via dynamical ejections and the shrinking tidal limits as the clusters approach SgrA*. In order to test the effect of mass segregation, we ran simulation {$\tt 1hi5\_ms$}, which we primordially mass segregated using the method described in \citet{Baumgardt08}. For simulations ${\tt 1hi5}$ and ${\tt 1hi5\_ms}$, Fig. \ref{mass_seg.fig} shows the semi-major axes of all stars and main sequence stars more massive than $8\unit{M\sscript{\odot}}$ at $T=7\unit{Myr}$, as well as the distributions of projected distances of spectroscopically visible stars at 7 and $15\unit{Myr}$. Their distributions look similar, as simulation ${\tt 1hi5}$ has an initial dynamical friction timescale of $t\sscript{df}\sim 0.1 \unit{Myr}$ for the most massive stars, causing the cluster to rapidly mass segregate. As such, primordial mass segregation does not significantly enhance the transport of massive stars to the central parsec, as clusters of high enough density become mass segregated before tides become important.

\begin{figure}
 \includegraphics[width=\linewidth]{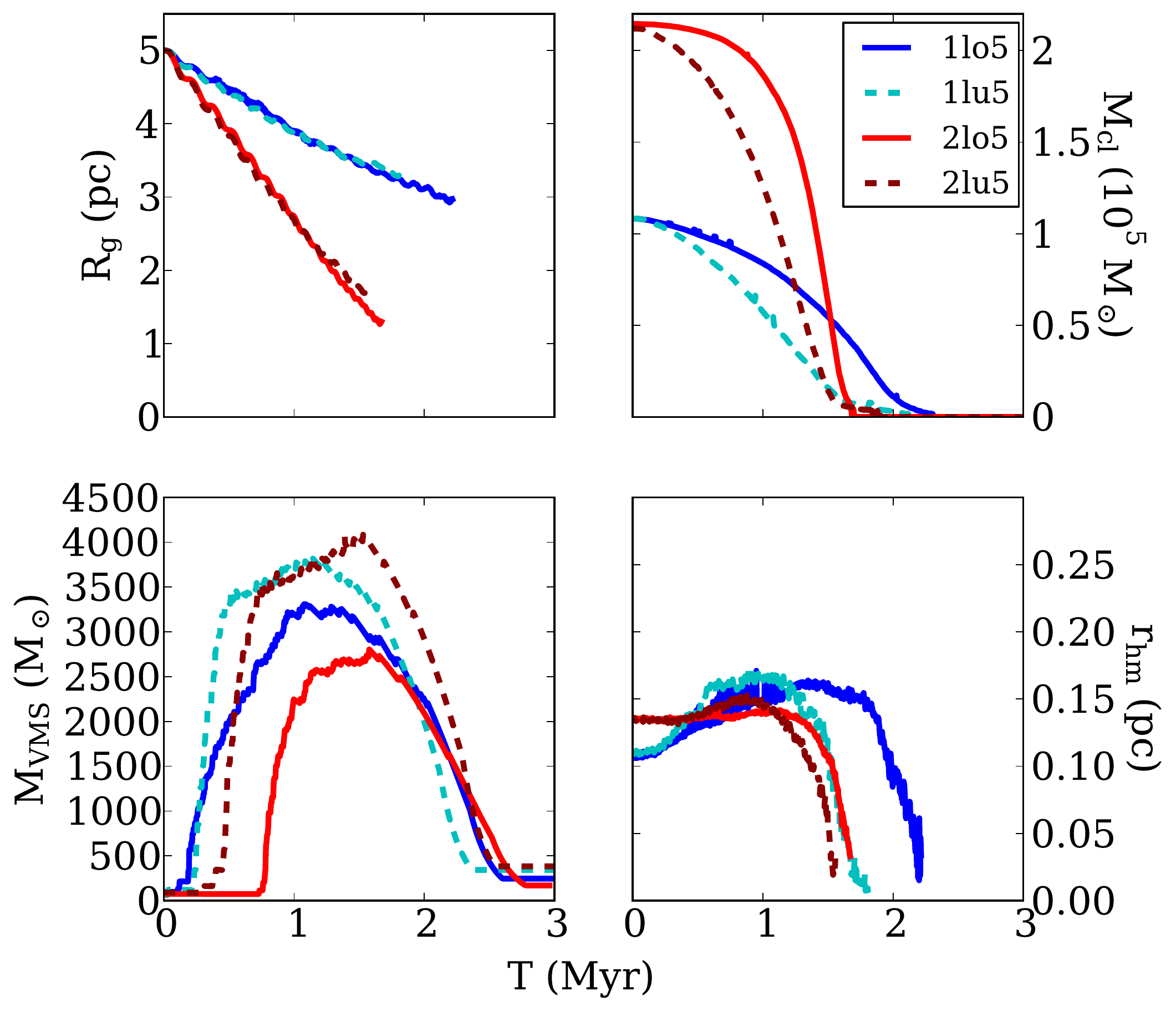}
 \caption{The evolution of galactocentric distance, cluster mass, VMS mass and cluster half mass radius as a function of time, for simulations {$\tt 1lo5$} (solid blue line), {$\tt 1lu5$} (dashed cyan line), {$\tt 2lo5$} (solid red line) and {$\tt 2lu5$} (dashed dark red line).} 
 \label{Lu.fig}
\end{figure}

\begin{figure}
 \includegraphics[width=\linewidth]{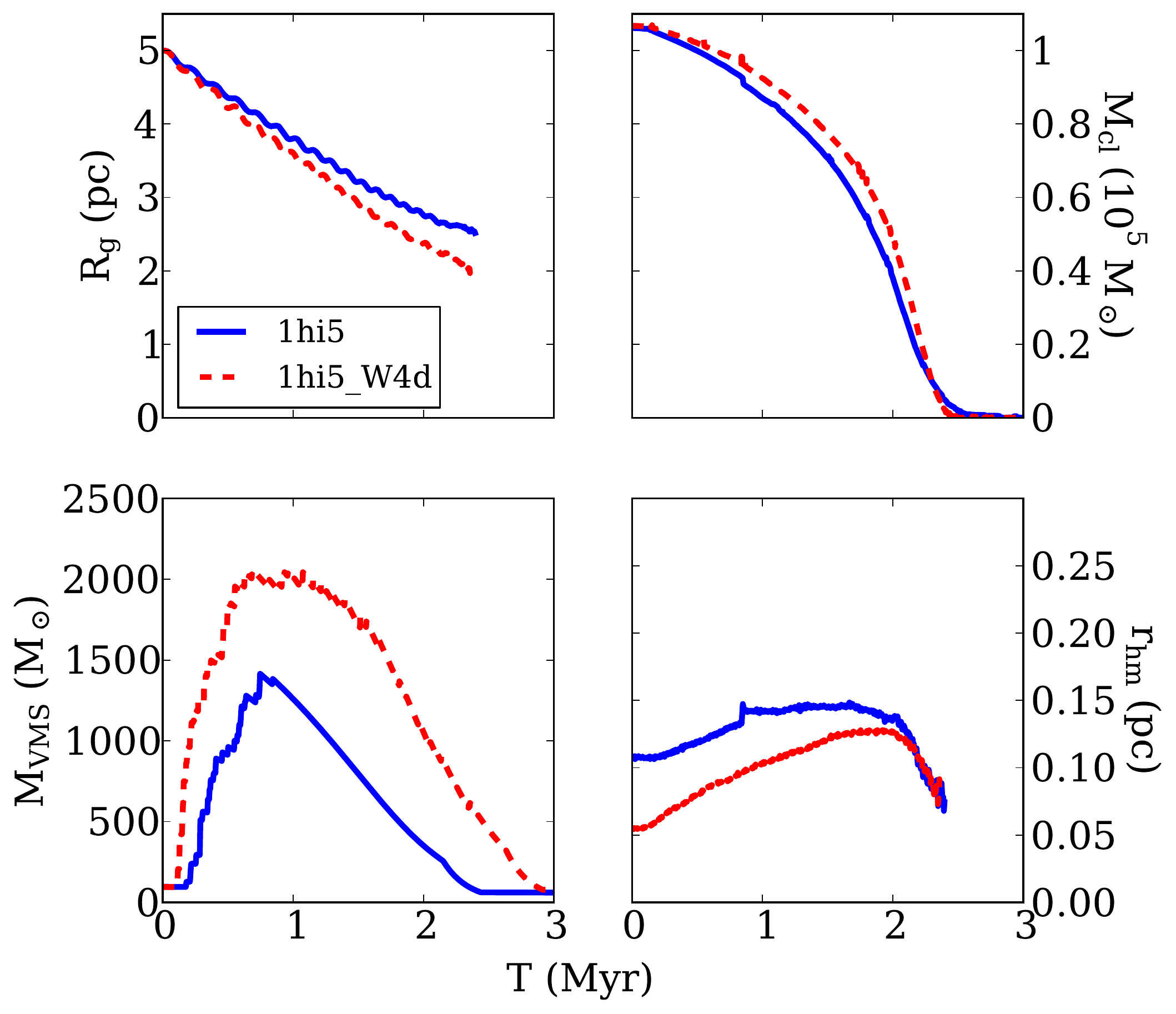}
 \caption{The evolution of galactocentric distance, cluster mass, VMS mass and cluster half mass radius as a function of time, for simulations {$\tt 1hi5$} (solid blue line) and {$\tt 1hi5\_W4d$} (dashed red line).} 
 \label{dense.fig}
\end{figure}

As star formation close to a SMBH is not well understood, we also test a model in which the cluster is born with the top heavy mass function derived in \citet{Lu13}. Fig. \ref{Lu.fig} shows the evolution of simulations {$\tt 1lo5$}, {$\tt 2lo5$}, {$\tt 1lu5$} and {$\tt 2lu5$}, where the two latter clusters have stars sampled from the \citet{Lu13} mass function. Models computed with the top-heavy mass function form VMSs of greater mass, as more massive stars are sampled, and their cross sections for collisions are larger. However, as the cluster mass is distributed amongst fewer stars, simulations {$\tt 1lu5$} and {$\tt 2lu5$} relax faster than {$\tt 1lo5$} and {$\tt 2lo5$} and dissolve more rapidly. A flatter mass function does not help bring stars closer to SgrA*, and leaves more visible stars spread across the central $10 \unit{pc}$.

As a final test of extreme initial conditions we re-run simulation {$\tt 1hi5$} with an initial size and density corresponding to being Roche filling at $1 \unit{pc}$. In simulation {$\tt 1hi5\_W4d$}, this cluster is placed initially on a circular orbit at $5 \unit{pc}$, so that it is initially very Roche under-filling. Fig. \ref{dense.fig} shows the evolution of Galactocentric distance, cluster mass, VMS mass and half mass radius of simulations {$\tt 1hi5$} and {$\tt 1hi5\_W4d$} as a function of time. Increasing the density shortens the relaxation time, and after $\sim 2 \unit{Myr}$ the clusters in simulations {$\tt 1hi5$} and {$\tt 1hi5\_W4d$} are similar. The cluster is able to hold onto its mass for slightly longer in {$\tt 1hi5\_W4d$}, but after the cluster expands it ultimately gets disrupted by the tidal field in the same way as {$\tt 1hi5$}. As such, making clusters arbitrarily dense does not help the cluster migration scenario.

\subsection{Models with primordial binaries}

\begin{figure}
 \includegraphics[width=\linewidth]{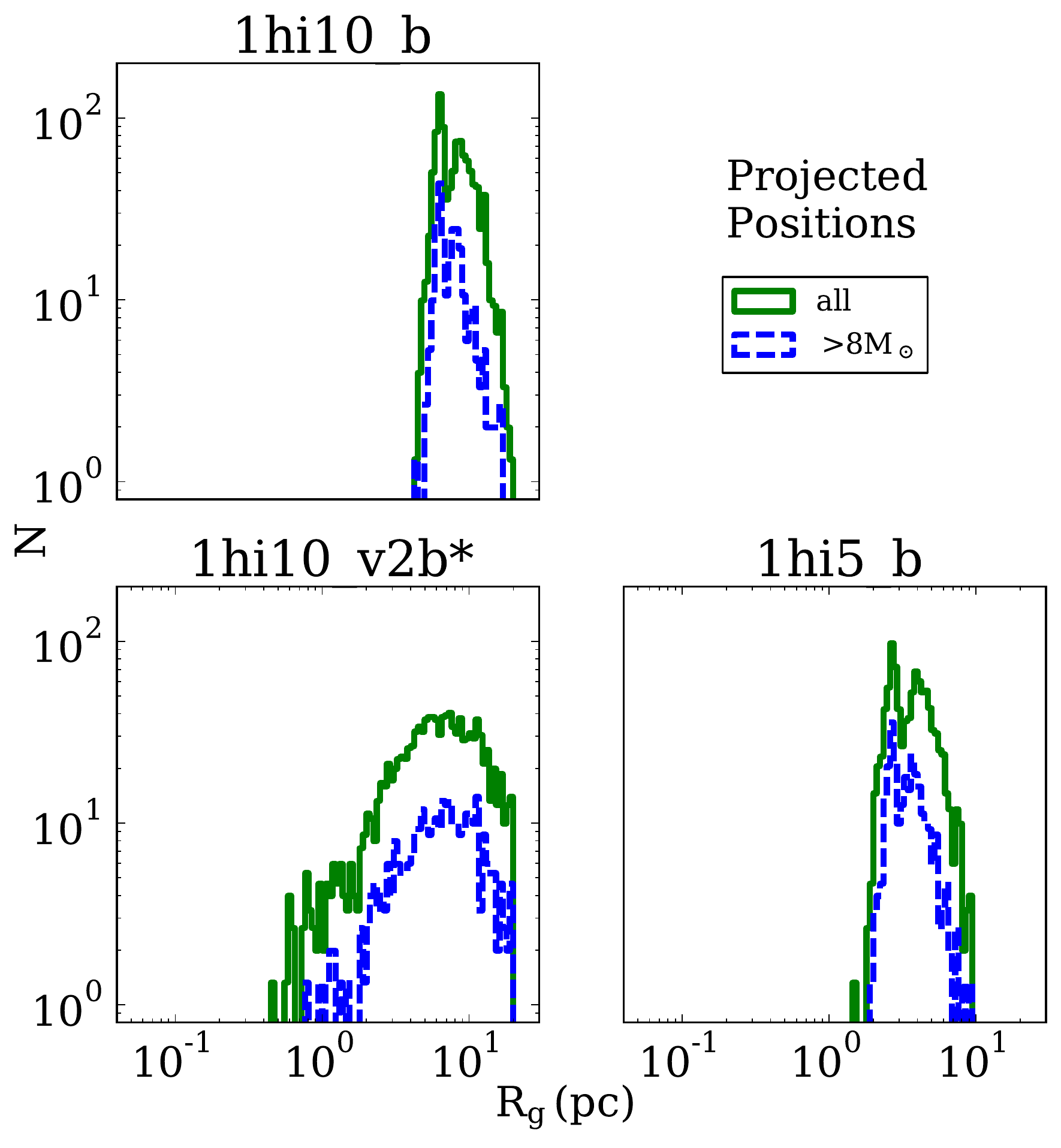}
 \caption{Final distributions of the projected distances of binaries from SgrA* in in simulations {$\tt 1hi10\_b$}, {$\tt 1hi10\_v2b*$} and {$\tt 1hi5\_b$}. The green histograms show the distributions of projected distances of all binaries remaining at $7 \unit{Myr}$, and the dashed blue histograms show those which have a main sequence primary more massive than $8\unit{M\sscript{\odot}}$ at $T=7\unit{Myr}$. The stars are projected so that the disk rotates clockwise in the sky. The y-values are re-normalised to the expected number of stars had the model been simulated with a full Kroupa IMF.} 
 \label{dist_bin.fig}
\end{figure}

We include a primordial binary population in three of our simulations ${\tt 1hi10\_b}$, ${\tt 1hi10\_v2b*}$ and ${\tt 1hi5\_b}$. The inclusion of primordial binaries is interesting as the clockwise disk has three confirmed eclipsing binaries \citep{Ott99,Martins06,Pfuhl14}, with a total binary population estimated to be greater than $30\%$ \citep{Pfuhl14, Gautam16}. Secondly, a popular formation scenario for the S-stars is from the tidal disruption of binaries by SgrA* via the Hills mechanism \citep{Hills91,Hills92}, where one star is captured  and the other is ejected as a hyper-velocity star.

Fig \ref{dist_bin.fig} shows the final projected distances of binaries in simulations ${\tt 1hi10\_b}$, ${\tt 1hi10\_v2b*}$ and ${\tt 1hi5\_b}$. In these models $5\%$ of the stars are initially in binary systems, the properties of which are described in section \S \ref{binarity.ch}. A large number of binary systems survive, despite some being consumed during the formation of the VMS. The final distributions of binaries with main sequence primaries more massive than $8 \unit{M\sscript{\odot}}$ are very similar to the distribution of single stars more massive than $8 \unit{M\sscript{\odot}}$ in models with no primordial binaries.

\begin{figure}
 \includegraphics[width=\linewidth]{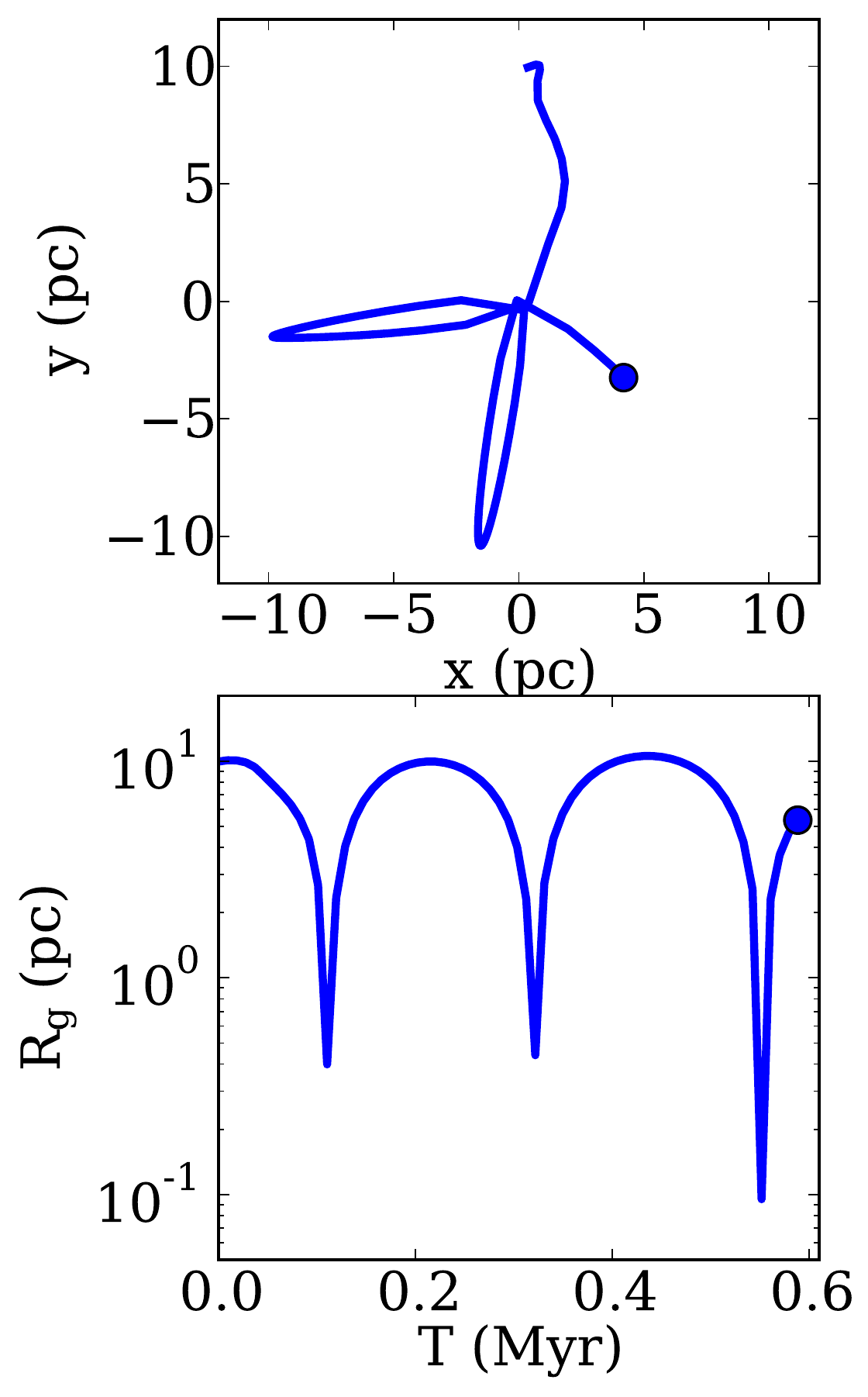}
 \caption{The orbit of the only binary to reach less than $0.1 \unit{pc}$ in simulation {$\tt 1hi10_v2b*$}. The top panel shows the orbit of the star in the x-y plane of the galactocentric rest frame. The x-y plane is defined such that the infalling cluster orbits clockwise, with SgrA* at the origin. The bottom panel shows the separation between SgrA* and the binary as a function of time. In both subplots the solid blue circle shows when the binary collides, forming a blue straggler star.} 
 \label{closest_bin.fig}
\end{figure}

In all three simulations, no binaries end up with semi-major axes less than $1\unit{pc}$. In ${\tt 1hi10\_v2b*}$ one massive binary of total mass $68.9 \unit{M\sscript{\odot}}$ came within $ 0.1 \unit{pc}$ of SgrA*. Fig. \ref{closest_bin.fig} shows the orbit of this star, which came $0.09 \unit{pc}$ from SgrA* at its third pericentre passage. However, the binary remained bound, as its tidal disruption radius by SgrA* was equal to $\sim 10 \unit{AU}$, $\sim 2000$ times smaller than its distance. The binary coalesced at apocentre. Due to the scarcity of binaries that approach SgrA*, and the fact that many binaries would be observed beyond the disk, we conclude that if the binary breakup scenario is the origin of the S-stars, it is unlikely that the progenitors originated from nearby star clusters.

\section{Discussion}

\label{discussion.ch}

The formation and evolution of a VMS appears to have little effect on cluster inspiral as compared with the collisionless models of \citet{Kim03}, yet the suppression of IMBH formation strongly inhibits the radial migration of a sub population of massive stars towards the central parsec (F09). However, even in the case of IMBH formation, one would still observe a broad distribution of massive stars out to $\sim 10 \unit{pc}$, making the scenario unlikely even if IMBH formation were efficient.

All simulations form a large disk of massive stars from $\sim 1 - 10\unit{pc}$, contradicting observations. This implies that no cluster has been present in this region in the past $\sim 15 \unit{Myr}$, as a large population would still be visible with current telescopes. Two $\sim 10^5 M\sscript{\odot}$ gas clouds, M-0.02-0.07 and M-0.13-0.08, are seen projected at $\sim 7$ and $\sim 13 \unit{pc}$ from SgrA* \citep{Solomon72}, suggesting the presence of GMCs in this region is commonplace. However, the absence of young stars in this region suggests that perhaps GMC collapse is suppressed, only triggering from a tidal shock at close passage to SgrA* \citep{Mapelli12}. Verifying this hypothesis is beyond the scope of this work, and would require further study of GMC collapse in the close vicinity of a massive black hole.

\section{Conclusions}

\label{Conclusions.ch}

We ran $N$-body simulations of young dense star clusters that form at distances of $5-15 \unit{pc}$ from SgrA* and inspiral towards the Galactic Centre due to dynamical friction. Most models are dense enough that runaway collisions are inevitable, forming a very massive star in less than $1 \unit{Myr}$. However, careful treatment of the evolution of this very massive star shows that it is likely to lose most of its mass through its stellar wind and end its life as a $\sim 50-250 M\sscript{\odot}$ black hole. As no intermediate mass black hole can form in this model, clusters dissolve a few $\unit{pc}$ from SgrA*, leaving a population of bright early type stars that would be observable for longer than the age of both the clockwise disk and S-star population, contradicting observations. It is therefore unlikely that a cluster has inhabited the central $10 \unit{pc}$ in the last $\sim 15 \unit{Myr}$, as such the S-stars are unlikely to have formed via disrupted binaries originating from star clusters. Instead, the clockwise disk likely formed in-situ, perhaps from a gas cloud on a radial orbit incident on SgrA* \citep{Mapelli12}, and the S-star cluster is likely to be populated either by dynamical processes in the clockwise disk \citep{Subr16}, or through the binary breakup of scattered binaries \citep{Perets09}.

\section*{Acknowledgements}
JAP would like to thank the University of Surrey for the computational resources required to perform the simulations used in this paper. 

\bibliography{PG16}
\bibliographystyle{mn2e,natbib}

\appendix
\renewcommand{\theequation}{A\arabic{equation}}
\renewcommand{\thefigure}{A\arabic{figure}}
\section*{appendix A: Dynamical Friction Comparison with GADGET}

\label{Appendix.ch}
\begin{figure}
 \includegraphics[width=\linewidth]{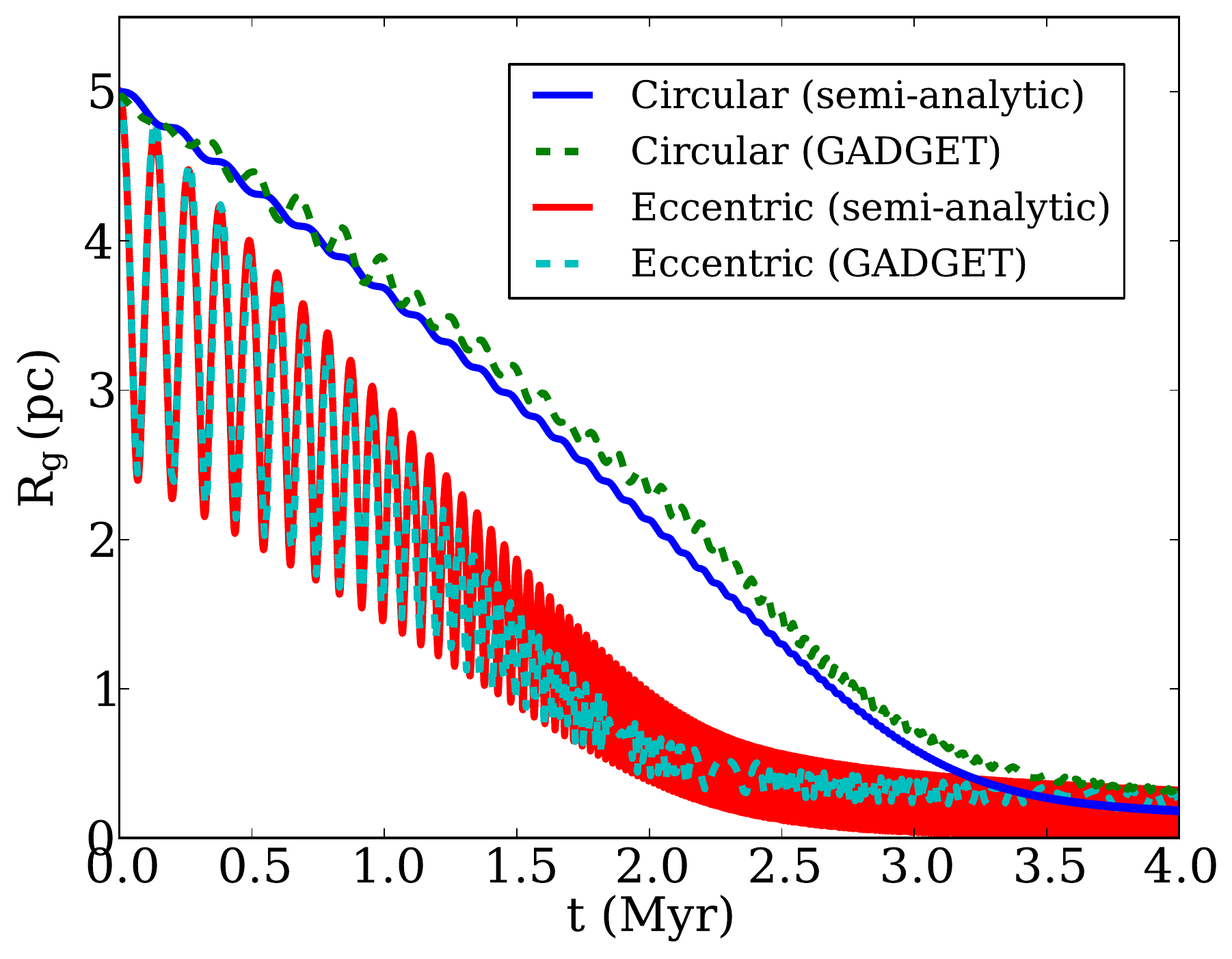}
 \caption{Orbital evolution of a $10^5M\sscript{\odot}$ point mass cluster in the Galactic Centre potential described in section \S \ref{IC.ch}, on circular and eccentric orbits in GADGET and using the semi-analytic model from \citet{Petts15}.} 
 \label{BH_test.fig}
\end{figure}

In \citet{Petts15}, we only tested our dynamical friction formulation against $N$-body models of single component Dehnen profiles. In this paper we add an additional central massive black hole. In the Maxwellian approximation, valid for cuspy distributions, this comes into Chandrasekhar's formula via the black hole's contribution to the velocity dispersion of the stars. The addition of a black hole to the model is described in the appendix of \citet{Petts15}.

Here we briefly show that our dynamical friction formulation in the vicinity of a black hole is accurate by means of two $N$-body models of point mass clusters, of mass $10^5M\sscript{\odot}$, orbiting the potential described in section \ref{IC.ch}. The $N$-body models are computer using the mpi-parallel tree-code GADGET2 \citep{Springel01}. The stellar background is comprised of $2^{24}$ particles of mass $3.5M\sscript{\odot}$, with a central black hole of $4.3\times10^6M\sscript{\odot}$. The softening of the cluster potential is $\epsilon = 0.0769\unit{pc}$, corresponding to $r\sscript{hm} \sim 0.1\unit{pc}$. The same softening length is used for the background particles. The black hole is given a softening length, $\epsilon = 0.2\unit{pc}$ to reduce numerical inaccuracies resulting from the large mass ratio of the black hole to background particles. In GADGET2 the force is exactly Newtonian at $2.8\epsilon$, so the semi-analytic and $N$-body models should agree to $\sim0.56\unit{pc}$. The cluster is initially $5\unit{pc}$ from SgrA*.

Fig. \ref{BH_test.fig} shows the orbital evolution of the circular and eccentric cases computed semi-analytically and with GADGET2. Our formalism agrees very well with the $N$-body models in both cases. In the eccentric case the circularisation appears to be slightly under-predicted. This is likely because we neglect the drag force from stars moving faster than the cluster \citep[see][]{Petts16}.

\label{lastpage}

\end{document}